\documentclass[manuscript]{aastex61}
\usepackage{amsmath}
\usepackage{placeins}
\usepackage{comment}
\usepackage[utf8]{inputenc}
\newcommand{\RN}[1]{%
  \textup{\uppercase\expandafter{\romannumeral#1}}%
}

\newcommand\aastex{AAS\TeX}

\received{March 5, 2018}
\revised{April 27, 2018}
\accepted{May 14, 2018}
\submitjournal{ApJ}

\shorttitle{\aastex\ Cosmic ray-driven grain chemistry}
\shortauthors{Shingledecker et al.}

\begin{document}

\title{On Cosmic Ray-Driven Grain Chemistry in Cold Core Models}

\correspondingauthor{Christopher N. Shingledecker}
\email{shingledecker@virginia.edu}

\author[0000-0002-5171-7568]{Christopher N. Shingledecker}
\affil{Department of Chemistry \\
University of Virginia \\
Charlottesville, VA 22904, USA}

\author{Jessica Tennis}
\affiliation{Department of Chemistry \\
University of Virginia \\
Charlottesville, VA 22904, USA}

\author{Romane Le Gal}
\affiliation{Department of Chemistry \\
University of Virginia \\
Charlottesville, VA 22904, USA}
\affiliation{Harvard-Smithsonian Center for Astrophysics \\
Cambridge, MA 02138, USA}

\author{Eric Herbst}
\affiliation{Department of Chemistry \\
University of Virginia \\
Charlottesville, VA 22904, USA}
\affiliation{Department of Astronomy \\
University of Virginia \\
Charlottesville, VA 22904, USA}


\begin{abstract}
  In this paper, we present preliminary results illustrating the effect of cosmic rays on
  solid-phase chemistry in models of both TMC-1 and several sources with physical conditions identical to TMC-1 except for hypothetically enhanced ionization rates. Using a recent
  theory for the addition of cosmic ray-induced reactions to astrochemical
  models, we calculated the radiochemical yields, called $G$ values, for the primary dust grain
  ice-mantle constituents.  We show that the inclusion of this non-thermal
  chemistry can lead to the formation of complex organic molecules from 
  simpler ice-mantle constituents, even under cold core conditions. In addition to enriching ice-mantles, we
  find that these new radiation-chemical processes can lead to increased
  gas-phase abundances as well, particularly for HOCO, NO$_2$, HC$_2$O, 
  methyl formate (HCOOCH$_3$), and ethanol (CH$_3$CH$_2$OH).  These model results imply that HOCO - and perhaps NO$_2$ -
  might be observable in TMC-1. Future detections of either of these two
  species in cold interstellar environments could provide strong support for
  the importance of cosmic ray-driven radiation chemistry. The increased gas-phase abundance of methyl formate can be compared with abundances achieved through other formation mechanisms such as pure gas-phase chemistry and three-body surface reactions.
\end{abstract}


\section{Introduction} \label{sec:intro}

Cosmic rays are a form of high-energy (MeV - TeV) ionizing radiation composed mostly of protons thought to form both in supernovae and galactic
nuclei \citep{blasi_origin_2013,baade_cosmic_1934,lemaitre_comptons_1933}. 
It has long been speculated that these
energetic particles can have significant physicochemical effects on the
interstellar medium (ISM) as a result of collisional energy transfer to the
matter in a region.  For example, in \citet{herbst_formation_1973}, cosmic rays
were shown to be the drivers of cold core chemistry via

\begin{equation}
  \mathrm{H_2} \leadsto \mathrm{H_2^+} + \mathrm{e^-}
\end{equation}
followed by 
\begin{equation}
  \mathrm{H_2^+} + \mathrm{H_2} \rightarrow \mathrm{H_3^+} + \mathrm{H}
\end{equation}

\noindent
where the curly arrow implies bombardment by an energetic particle. The
ion-molecule reactions initiated by H$_3^+$ are of central importance in the
subsequent formation of polyatomic species. In addition, cosmic rays are
thought to play an important role both in source heating
\citep{goldsmith_molecular_1978,ao_thermal_2013} and in generating internal UV
photons in cold cores through the Lyman and Werner band excitation of H$_2$
\citep{prasad_uv_1983}. 

The Galactic value of the cosmic ray ionization rate, $\zeta$, cannot be
directly measured from Earth due to the effects of the Solar wind
\citep{parker_cosmic-ray_1958}. It is thought that the most common ionization 
rate in the ISM is $\zeta\approx10^{-15}$ s$^{-1}$ everywhere but in dense regions \citep{grenier_nine_2015},
where interactions between the dense cloud and the charged particles that comprise cosmic rays
result in a reduced ionization rate of $\sim 10^{-17}$ s$^{-1}$ \citep{rimmer_observing_2012}. 
However, even in dense regions, local effects can result in substantially higher
fluxes of ionizing radiation leading to ionization rates in the range $\zeta \approx 10^{-15} -
10^{-14}$ s$^{-1}$.   Such rates arise in Sgr A*
\citep{yusef-zadeh_interacting_2013,yusef-zadeh_widespread_2013,ao_thermal_2013},
and in sources like W51C, which are near supernova remnants
\citep{ceccarelli_supernova-enhanced_2011,shingledecker_inference_2016}.

Collisions between cosmic rays and dust grains are also important in the ISM. 
For instance, \citet{ivlev_interstellar_2015} note that cosmic rays affect the net charge on dust particles, 
which has an influence on grain growth. Cosmic ray collisions have also been implicated in
impulsive grain heating \citep{hasegawa_new_1993,ivlev_impulsive_2015}, which can stimulate both diffusive chemistry and desorption. Despite this, the direct chemical
effects resulting from cosmic ray bombardment of dust grain ice mantles are not
currently considered in astrochemical models.
Previous experimental work has
shown that the bombardment of low-temperature ices by ionizing radiation can
trigger a rich chemistry \citep{hudson_radiation_2001,rothard_modification_2017,abplanalp_study_2016} - including the formation of complex organic molecules such as amino acids \citep{hudson_amino_2008,lafosse_reactivity_2006,holtom_combined_2005}.

Following \citet{bohr_theory_1913}, the energy lost by an energetic particle per distance travelled -
called the stopping power - can be approximated by the sum of two types of 
energy loss, as seen in the following equation:

\begin{equation}
      \frac{dE}{dx} = n(S_\mathrm{n} + S_\mathrm{e})
\end{equation}

\noindent
where $n$ is the density of the target material, while 
$S_\mathrm{n}$ and $S_\mathrm{e}$ are so-called stopping cross sections \citep{johnson_energetic_1990,ziegler_stopping_1985} - also known as energy loss functions, in units of area $\times$ energy \citep{peterson_relation_1968}.
Here, $S_\mathrm{n}$ characterizes the elastic energy collisionally transferred to nuclei in a material, while
$S_\mathrm{e}$ characterizes the energy transferred to electrons in inelastic collisions \citep{bohr_theory_1913,johnson_energetic_1990,spinks_introduction_1990}.
Inelastic events, in turn, are typically approximated as consisting of collisions
that cause either the ionization or electronic excitation of target species. 
The ionization of species in a material results in the formation of  so-called ``secondary
electrons'' \citep{spinks_introduction_1990}. Around $10^4$ secondary electrons can be
produced per MeV transferred to a material, and they play a critical role in
propagating physicochemical changes initiated by primary ions
\citep{gerakines_energetic_2001,mason_electron_2014,spinks_introduction_1990}.

In \citet{abplanalp_study_2016}, we made the first attempt - to the best of our
knowledge - to incorporate experimentally determined chemical reactions resulting from radiation processes
 into an astrochemical model. Based on insights gained both from that
work, and from radiation chemistry based on a subsequent detailed microscopic Monte Carlo 
model \citep{shingledecker_new_2017}, we developed a general method described
in detail in \citet{shingledecker_general_2017} 
targeted at the great majority of astrochemically relevant radiolysis processes
which have not been studied in detail in the laboratory. 
The basis of this method is that a microscopic collision between
a target species, $A$, and either a primary ion or secondary electron is assumed to
have one of the following outcomes:

\begin{equation}
A \leadsto A^+ + e^-
\tag{R1}
\label{p1}
\end{equation}

\begin{equation}
A \leadsto A^+ + e^- \rightarrow A^* \rightarrow bB^* + cC^*
\tag{R2}
\label{p2}
\end{equation}

\begin{equation}
A \leadsto A^* \rightarrow bB + cC
\tag{R3}
\label{p3}
\end{equation}

\begin{equation}
A \leadsto A^*
\tag{R4}
\label{p4}.
\end{equation}

\noindent
Here, the asterisk indicates an electronically excited species, which can
be referred to as ``suprathermal'' \citep{abplanalp_study_2016};  $B$ and $C$ are the dissociation products;
and the lowercase letters are the stoichiometric coefficients \citep{spinks_introduction_1990}.  In this work,
we will refer to molecular dissociation due to bombardment by ionizing
radiation as {\it radiolysis} \citep{spinks_introduction_1990,johnson_photolysis_2011}.

In processes \eqref{p1} and \eqref{p2}, $A$ is ionized upon collision with an
energetic particle, resulting in the ion-pair $A^+ + e^-$, which can
quickly undergo dissociative recombination, as shown in \eqref{p2}. The relative
importance of \eqref{p1} and \eqref{p2} is characterized by the electron escape
probability, $P_\mathrm{e}$, which we will here assume to be zero for
solid-phase processes, so that \eqref{p1} is negligible. In processes
\eqref{p3} and \eqref{p4}, $A$ is electronically excited after collision with
an energetic particle. As with the ionizing processes, \eqref{p1} and
\eqref{p2}, the relative importance of \eqref{p3} and \eqref{p4} is given by
$P_\mathrm{dis}$, the dissociation probability, which we will here assume to be
0.5 in the absence of relevant experimental or theoretical values. 
Based on results from previous, more detailed Monte Carlo modeling of radiation chemistry \citep{shingledecker_new_2017}, we have assumed that the intermediate species A$^*$ produced via R2 dissociates immediately with unit probability, unlike in process R3, due to the greater exothermicity of dissociative recombination.

The suprathermal species produced in processes \eqref{p2} and \eqref{p4} are
critical when considering the  effects of radiation exposure on a
material, particularly in cold regions, because
their energies are often sufficient to overcome reaction barriers that are
inaccessible to the reactants in their ground electronic states
\citep{spinks_introduction_1990}. Previous experimental work suggests that these
electronically excited species can drive the formation of complex organic
molecules, even in solids at 5 K \citep{abplanalp_study_2016}, where they
likely either rapidly react with a neighbor or are quenched by the material
\citep{spinks_introduction_1990}. 

The overall efficiency of processes \eqref{p1}-\eqref{p4}, called the radiochemical
yield, is characterized by the $G$ value \citep{dewhurst_general_1952}, defined as the number of molecules created
or destroyed per 100 eV deposited by an incident energetic particle into some system.
As described in detail in \citet{shingledecker_general_2017}, the $G$ values
for processes \eqref{p1}-\eqref{p4} can be calculated using the following
expressions: 

\begin{equation}
  G_\mathrm{R1} = P_\mathrm{e}\left(\frac{100\,\mathrm{eV}}{W}\right)
	\label{g1}
\end{equation}

\begin{equation}
  G_\mathrm{R2} = (1-P_\mathrm{e})\left(\frac{100\,\mathrm{eV}}{W}\right)
	\label{g2}
\end{equation}

\begin{equation}
  G_\mathrm{R3} = P_\mathrm{dis}\left(\frac{100\,\mathrm{eV}}{W}\right)\left(\frac{W-(E_\mathrm{ion}+W_\mathrm{s})}{W_\mathrm{exc}}\right)
	\label{g3}
\end{equation}

\begin{equation}
  G_\mathrm{R4} = (1-P_\mathrm{dis})\left(\frac{100\,\mathrm{eV}}{W}\right)\left(\frac{W-(E_\mathrm{ion}+W_\mathrm{s})}{W_\mathrm{exc}}\right)
	\label{g4}
\end{equation}

\noindent
where $W$ is the mean energy per ion-pair (usually $\sim 30$ eV) \citep{dalgarno_energy_1958,edgar_energy_1973},
$E_\mathrm{ion}$ is the ionization energy of $A$, $W_\mathrm{exc}$ is the
average excitation energy of $A$, and $W_\mathrm{s}$ is the average sub-excitation energy of the secondary electrons formed via the ionization of $A$
(typically $\sim 3$ eV) \citep{fueki_reactions_1963,elkomoss_parention_1962}. 

By definition, there is one ionization per ion-pair; however, 
the number of excitations per ionization is a function of the average excitation 
energy. The average number of excitations per ionization, $\xi$, is given by

\begin{equation}
    \xi = \frac{W-(E_\mathrm{ion}+W_\mathrm{s})}{W_\mathrm{exc}}
\end{equation}

\noindent
and is the extra factor included in Eqs. \eqref{g3} and \eqref{g4}. Physically,
for every $W$ eV lost per ion-pair, an amount equal to $E_\mathrm{ion}$ of that energy is used to
generate the ion-pair, and some small amount $W_\mathrm{s}$ accounts for the fact that
secondary electrons (a) lose energy through inelastic collisions or (b) have insufficient
energy upon formation to either ionize or excite species in the material. Thus, the
remaining energy per ion-pair available to cause electronic excitations is $W-(E_\mathrm{ion}+W_\mathrm{s})$,
and $\xi$, the average number of excitations that can result from this amount of energy, is a function of the
average excitation energy, $W_\mathrm{exc}$.

These $G$ values can, in turn, be used to estimate the first-order rate coefficients (s$^{-1}$) of
processes \ref{p1}-\ref{p4} via

\begin{equation}
	k_\mathrm{R1} = G_\mathrm{R1} \left(\frac{S_\mathrm{e}}{100\;\mathrm{eV}}\right) \left(\phi_\mathrm{ST}\left[\frac{\zeta}{10^{-17}}\right]\right)
   \label{k1}
\end{equation}

\begin{equation}
	k_\mathrm{R2} = G_\mathrm{R2} \left(\frac{S_\mathrm{e}}{100\;\mathrm{eV}}\right) \left(\phi_\mathrm{ST}\left[\frac{\zeta}{10^{-17}}\right]\right)
   \label{k2}
\end{equation}

\begin{equation}
	k_\mathrm{R3} = G_\mathrm{R3} \left(\frac{S_\mathrm{e}}{100\;\mathrm{eV}}\right) \left(\phi_\mathrm{ST}\left[\frac{\zeta}{10^{-17}}\right]\right)
   \label{k3}
\end{equation}

\begin{equation}
k_\mathrm{R4} = G_\mathrm{R4} \left(\frac{S_\mathrm{e}}{100\;\mathrm{eV}}\right) \left(\phi_\mathrm{ST}\left[\frac{\zeta}{10^{-17}}\right]\right).
   \label{k4}
\end{equation}

\noindent
Here, $\phi_\mathrm{ST}$ is the integrated Spitzer-Tomasko cosmic ray flux (8.6
particles cm$^{-2}$ s$^{-1}$) \citep{spitzer_heating_1968}, $\zeta$ is the
H$_2$ ionization rate, and $S_\mathrm{e}$ is the electronic stopping cross
section 
\citep{bethe_bremsformel_1932,johnson_energetic_1990,ziegler_stopping_1985}.  
Amorphous H$_2$O is typically the dominant ice-mantle constituent;
thus, we approximate the stopping cross section for protons in amorphous water
ice with the more readily available values for liquid water, which were
calculated using the \texttt{PSTAR} program\footnote{https://physics.nist.gov/PhysRefData/Star/Text/PSTAR.html}. An average value of $S_\mathrm{e} = 1.287 \times 10^{-15}$ cm$^{2}$ eV
was obtained using the Spitzer-Tomasko cosmic ray flux  \citep{spitzer_heating_1968}. 
One can estimate the effect of going from a water ice to, for example, one comprised mainly of CO using the ratio of stopping cross sections for the two species. Using the Bethe equation for the electronic stopping cross section \citep{bethe_bremsformel_1932} - and all else being equal - $S_e^{CO} \approx 1.4\times S_e^{H2O}$ - which may not have a significant effect in most astrochemical models.
As discussed further in \S \ref{sec:network}, when multiplied by the density of the reactant species, Eqs. \eqref{k1}-\eqref{k4} refer to the time dependence of the concentration of products produced by radiolysis - driven mainly by inelastic collisions involving secondary electrons.

We here examine how radiolysis of the primary
dust grain ice mantle constituents influences the chemistry of cold cores like
TMC-1. 
The organization of the rest of this paper is as follows: in \S \ref{sec:model} we give
details concerning the code and physical conditions used here, while \S \ref{sec:network}
contains a description of the reactions and processes added to the network for this work. \S
\ref{sec:results} concerns the description and discussion of our major findings,
while in \S \ref{sec:conclusion}, we summarize our results and point to areas of
future development.

\section{Model} \label{sec:model}

In this work, we focus on the chemistry of cold cores, such as TMC-1.  Despite
the low temperatures of these regions, their chemical complexity has been
highlighted by recent detections of species such as HC$_5$O
\citep{mcguire_detection_2017}, HC$_7$O
\citep{mcguire_detection_2017,cordiner_deep_2017}, and the aromatic molecule
benzonitrile \citep{mcguire_detection_2018}.  The effects of radiation
chemistry should be more pronounced in these cold interstellar environments
since thermal diffusion is inhibited, thus increasing the relative importance
of fast solid-phase reactions involving suprathermal species.

We utilized the \texttt{NAUTILUS}-1.1 astrochemical model
\citep{ruaud_gas_2016}, in which three phases are simulated, specifically, (a) the
gas-phase, (b) the ice/grain-surface, and (c) the ice-mantle bulk.  This
distinction between the surface and bulk of the ice is helpful here, since it
highlights an important aspect of solid-phase radiation chemistry, namely, that
bombardment by ionizing radiation can greatly increase the chemical importance
of the bulk ice, since this is the phase in which the majority of the
physicochemical changes likely occur \citep{johnson_energetic_1990,spinks_introduction_1990,shingledecker_new_2017}. 
The degree of penetration into the ice constitutes a  major difference between photochemistry and radiation chemistry  \citep{gerakines_energetic_2001,gerakines_ultraviolet_2004}.
In the
absence of bombardment by energetic particles, the surface is significantly
more important in astrochemical models, due both to the lower diffusion
barriers and direct contact with the surrounding gas.  The non-thermal
desorption mechanisms for surface species are (1) chemical desorption with a
standard 1\% efficiency \citep{garrod_non-thermal_2007} (2) cosmic ray-induced
desorption \citep{hasegawa_new_1993}, and (3) photodesorption \citep{bertin_indirect_2013}. 

We ran simulations of two different types of sources, the cold core TMC-1 and a
group of hypothetical sources physically identical to TMC-1, other than
having higher ionization rates. The latter set of simulations were run in order
to identify any trends in our models arising from the included radiation
chemistry. The physical conditions used here for both sets of simulations are
given in Table 1, and all models utilized the same initial
elemental abundances, listed in Table 2.

\begin{deluxetable}{lcccc}
\label{tab:parameters}
\tablecaption{Model parameters and physical conditions used.}
\tablehead{
\colhead{Parameter} & \colhead{} & \colhead{} & \colhead{TMC-1} & \colhead{Hypothetical Sources} 
}
\startdata
$n_\mathrm{H}$ (cm$^{-3}$) & & & $10^4$ & $10^4$ \\
$n_\mathrm{dust}$ (cm$^{-3}$) & & & $1.8\times 10^{-8}$ & $1.8\times10^{-8}$ \\
$T_\mathrm{gas}$ (K) & & & 10 &  10 \\
$T_\mathrm{grain}$ (K) & & & 10 & 10 \\
$A_\mathrm{v}$ (mag) & & & 10 & 10 \\
$N_\mathrm{site}$ (cm$^{-2}$) & & & $1.5\times10^{15}$ & $1.5\times 10^{15}$ \\
$\zeta$ (s$^{-1}$) & & & $10^{-17}$ & $10^{-17}-10^{-14}$ \\
\enddata
\end{deluxetable}

\begin{deluxetable}{lccccc}
\label{tab:abundances}
\tablecaption{Elemental abundances used in this work.}
\tablehead{
\colhead{Element} & \colhead{} & \colhead{} & \colhead{} & \colhead{} & \colhead{Value}
}
\startdata
$X$(H$_2$) & & & & & $5.00\times 10^{-1}$ \\ 
$X$(He)\tablenotemark{a} & & & & & $9.00 \times 10^{-2}$ \\ 
$X$(N)\tablenotemark{a} & & & & & $2.14 \times 10^{-5}$ \\ 
$X$(O)\tablenotemark{b} & & & & & $1.70 \times 10^{-4}$ \\ 
$X$(C$^+$)\tablenotemark{c} & & & & & $1.70 \times 10^{-4}$ \\ 
$X$(S$^+$)\tablenotemark{d} & & & & & $8.00 \times 10^{-8}$ \\ 
$X$(Si$^+$)\tablenotemark{d} & & & & & $8.00 \times 10^{-9}$ \\ 
$X$(Fe$^+$)\tablenotemark{d} & & & & & $3.00 \times 10^{-9}$ \\ 
$X$(Na$^+$)\tablenotemark{d} & & & & & $2.00 \times 10^{-9}$ \\ 
$X$(Mg$^+$)\tablenotemark{d} & & & & & $7.00 \times 10^{-9}$ \\ 
$X$(P$^+$)\tablenotemark{d} & & & & & $2.00 \times 10^{-10}$ \\ 
$X$(Cl$^+$)\tablenotemark{d} & & & & & $1.00 \times 10^{-9}$ \\ 
$X$(F)\tablenotemark{e} & & & & & $6.68 \times 10^{-9}$ \\ 
\enddata
\tablenotetext{a}{\citet{wakelam_polycyclic_2008}}
\tablenotetext{b}{\citet{mcguire_detection_2018}}
\tablenotetext{c}{\citet{jenkins_unified_2009}}
\tablenotetext{d}{\citet{graedel_kinetic_1982}}
\tablenotetext{e}{\citet{neufeld_chemistry_2005}}
\end{deluxetable}

\FloatBarrier

\section{Network} \label{sec:network}

\startlongtable
\begin{deluxetable}{lccrrr}
  \label{tab:radinputs}
  \tablecaption{Parameters used in calculating $G$ values and rate coefficients.}
  \tablehead{
    \colhead{Species} & \colhead{} & \colhead{} & \colhead{$E_\mathrm{ion}$\tablenotemark{a}} & \colhead{$W_\mathrm{exc}$\tablenotemark{b}} & \colhead{$W_\mathrm{s}$}  \\
  \colhead{} & \colhead{} & \colhead{} & (eV) & (eV) & (eV) 
}
  \startdata
  H$_2$O         & & & 12.621 & 11.190 & 3.824 \\
  O$_2$          & & & 12.070 & 8.500  & 3.886 \\
  O$_3$          & & & 12.530 & 4.860  & 3.815 \\
  CO             & & & 14.014 & 13.190 & 3.947 \\
  CO$_2$         & & & 13.777 & 13.776 & 3.927 \\
  NO             & & & 9.264  & 13.776 & 3.422 \\
  NO$_2$         & & & 9.586  & 21.377 & 3.478 \\
  O$_2$H         & & & 11.350 & 5.961  & 3.694 \\
  H$_2$O$_2$     & & & 10.580 & 10.332 & 3.606 \\
  CH$_3$OH       & & & 10.840 & 14.760 & 3.636 \\
  NH$_3$         & & & 10.070 & 9.110  & 3.542 \\
  H$_2$CO        & & & 10.880 & 7.940  & 3.641 \\
  CH$_4$         & & & 12.610 & 13.000 & 3.823 \\
  CH$_3$COCH$_3$ & & & 9.703  & 6.358  & 3.494 \\
  \enddata
  \tablenotetext{a}{\citet{lias_ionization_2018}}
  \tablenotetext{b}{\citet{keller-rudek_mpi-mainz_2013}}
\end{deluxetable}

Our three-phase chemical network is based on the one described in
\citet{ruaud_gas_2016} to which we have added the gas-phase reactions of
\citet{balucani_formation_2015}.  To this network, we have included
both (a) dissociation pathways for the major ice mantle constituents due to collisions with cosmic rays or secondary electrons and (b) reactions involving the suprathermal products. Radiochemical yields ($G$
values) and rate coefficients were calculated using the Shingledecker-Herbst
method, and are a function of $E_\mathrm{ion}$, $W_\mathrm{exc}$, and $W_\mathrm{s}$.
Values for the ionization energy, $E_\mathrm{ion}$, were taken from the NIST Chemistry Webbook \citep{lias_ionization_2018}. The average electronic excitation energies, $W_\mathrm{exc}$, 
were estimated from the strongest UV-Vis absorption for each species \citep{fueki_reactions_1963,shingledecker_general_2017} based on 
spectra in the MPI-Mainz UV-Vis Spectral Atlas \citep{keller-rudek_mpi-mainz_2013}. Finally, 
the average sub-excitation electron energies were calculated using the method of \citet{elkomoss_parention_1962}. A list of both the species that undergo radiolysis as well as the
associated parameters used in calculating rate coefficients are given in Table
\ref{tab:radinputs}, while Table 4 in Appendix A lists the new solid-phase
radiolysis pathways for each species.

In our models, we assume the processes in Table 4 occur both on the surface and
in the ice mantle and have labeled them Types I, II, and III. Type I radiolysis
corresponds to the process given in equation \eqref{p2} where species $A$ is
ionized and recombines with the newly formed electron to produce suprathermal
dissociation products. Type II processes correspond to the  sequence
of events given in equation \eqref{p3}, where $A$ dissociates into thermal
products after being collisionally excited by an energetic particle.  Finally,
Type III processes are characterized by equation \eqref{p4}, where $A$ is
collisionally excited, but does not immediately dissociate.

As supported by previous experimental work \citep{bennett_laboratory_2005,abplanalp_study_2016,bergantini_mechanistical_2018}, we assume that for a suprathermal species $B^*$, the lifetime in solids is much shorter ($<< 1$ s) than the average surface or bulk thermal hopping time, $t_\mathrm{hop}^B$ ($>> 1$ s at 10 K) \citep{hasegawa_models_1992}. As noted by \citet{bennett_laboratory_2005}, the short lifetimes of these suprathermal species, relative to their hopping times at low temperatures, means that their solid-phase chemistry is likely dominated by reactions with neighbors. Therefore, we assume that once formed, suprathermal species only either react or relax back to the ground state. For reactions of the form

\begin{equation}
  A + B^* \rightarrow \text{products}.
  \label{r1}
\end{equation}

\noindent

we  use the following formula for calculating the rate coefficients, $k_\mathrm{ST}$(cm$^3$s$^{-1}$):

\begin{equation}
  k_\mathrm{ST} = 
  f_\mathrm{br}
  \left[
  \frac{
  \nu_0^B + 
  \nu_0^A
  }
  {
  {N_\mathrm{site}n_\mathrm{dust}}
  }
  \right]
  \label{ksup}
\end{equation}

\noindent
where $f_\mathrm{br}$ is the product branching fraction, $n_\mathrm{dust}$ is the dust density - here equal to $1.8\times10^{-8}$ cm$^{-3}$, $N_\mathrm{site}$ is the number of physisorption sites on the grain - here equal to $1.5\times10^{15}$ cm$^{-2}$, and $\nu_0^X$ is the characteristic vibrational frequency  for some physisorbed species, $X$, which is typically in the range of $1-3\times10^{-12}$ s$^{-1}$ \citep{herbst_chapter_2008}. This frequency can be estimated \citep{landau_mechanics_1976} using the formula

\begin{equation}
  \nu_{0}^{X} = \sqrt{
  \frac
  {2N_\mathrm{site}E_\mathrm{b}^X}
  {\pi^2m_X}
  }
  \label{trialnu}
\end{equation}

\noindent
where $m_X$ is the mass of $X$ and $E_\mathrm{b}^X$ is the diffusion barrier, which we here set equal to 40\% and 80\% - for surface and bulk species, respectively - of the desorption energies used in \citet{ruaud_gas_2016}. 
Since the dominant mechanism for reactions involving suprathermal species in solids is likely not diffusive \citep{bennett_laboratory_2005}, Eq. \eqref{ksup} is similar to the typical solid-phase bimolecular rate coefficients, but differs from them in that it does not contain either (a) a term characterizing thermal hopping or (b) a factor accounting for tunneling through reaction barriers, since we assume that suprathermal species are sufficiently energetic to react without a barrier \citep{hasegawa_models_1992}.

In addition to destruction through chemical reactions, we also assume that suprathermal species can be quenched by the ice-mantle \citep{spinks_introduction_1990,bennett_laboratory_2005}, i.e.

\begin{equation}
B^* + M \rightarrow B + M.
\end{equation}

\noindent
We use the characteristic frequency, $\nu_0^B$, as a pseudo first-order approximation for the rate coefficient of the above process.
Here, we have assumed that quenching by the solid is very fast ($\sim 10^{-14}$ s) compared to spontaneous emission ($\sim 10^{-9}$ s) and thus have neglected it as a de-excitation channel in this work.

To illustrate how this radiation chemistry is incorporated into our chemical 
network, consider the formation and destruction of the
suprathermal species, $B^*$, which is produced solely via process (R2) and only
reacts with $A$, as in Eq. \eqref{r1}. In this example then, the rate of change of $n(B^*)$ is given by the equation

\begin{equation}
    \frac{d\,n(B^*)}{d\,t} = k_\mathrm{R2}n(A) - \nu_0^Bn(B^*) - k_\mathrm{ST}n(A)n(B^*)
    \label{toymodel}
\end{equation}

\noindent
where the first term on the right gives the production of $B^*$ via the radiolysis of $A$,
the second term gives the quenching rate for $B^*$, and the third term
gives the rate of destruction via reaction with $A$ - with $k_\mathrm{ST}$ being the rate coefficient for suprathermal reactions given in Eq. \eqref{ksup}. We emphasize that in our actual network, most suprathermal species are produced from the radiolysis of more than one species, and all have more than one destructive reaction.

The suprathermal reactions we have added to our network can be grouped into two
classes. Class 1 refers to those that are similar to reactions
involving ground state species already included in the network, while Class 2 refers to
novel reactions unlike those currently included for thermal species.    To illustrate Class 1 reactions, consider the following example:

\begin{equation}
  \mathrm{H}(s) + \mathrm{CO}(s) \rightarrow \mathrm{HCO}(s)
  \label{h_co}
\end{equation}

\noindent
which has an activation energy of 2300 K in the \citet{ruaud_gas_2016} network, in addition to a diffusion barrier. Here,
$(s)$ indicates either a surface or bulk species.  We will later use
$(g)$ to denote gaseous species, and in cases where reactants labeled with $(s)$
lead to products in the gas-phase, the reactants are assumed to be surface
species only.
Here we include the following Class 1 suprathermal reactions based on
\eqref{h_co}:

\begin{equation}
  \mathrm{H^*}(s) + \mathrm{CO}(s) \rightarrow \mathrm{HCO}(s)
  \label{h*_co}
\end{equation}

\begin{equation}
  \mathrm{H}(s) + \mathrm{CO^*}(s) \rightarrow \mathrm{HCO}(s).
  \label{h_co*}
\end{equation}

\noindent
We assume no barrier for both reaction \eqref{h*_co} and \eqref{h_co*}, as
implied by results from ice irradiation experiments
\citep{abplanalp_study_2016}. Rate coefficients for reactions \eqref{h*_co} and
\eqref{h_co*}, as well as for all similar Class 1 suprathermal reactions,  are
calculated in our model using Eq. \eqref{ksup}. Another group of Class 1
reactions included in our network are based on work by
\citet{hudson_radiation_2017}, who found ketene (H$_2$CCO) among the products
of acetone irradiation, which could form via

\begin{equation}
  \mathrm{CH_3} + \mathrm{CH_3CO} \rightarrow \mathrm{H_2CCO} + \mathrm{CH_4}
\end{equation}

\noindent
where the CH$_3$ and CH$_3$CO radicals result from either Type I or II
radiolysis of acetone. We have included both the reaction between ground
state radicals as well as reactions involving a single suprathermal reactant,
similar to reactions \eqref{h*_co} and \eqref{h_co*}. A full list of these new reactions is available from the authors.

Class 2 is used to categorize novel reactions that are unlike the kinds of
thermal reactions typically considered in gas/grain models. To illustrate why
this type of chemistry is astrochemically interesting, consider the following
Class 2 reaction:

\begin{equation}
  \mathrm{O}(s) + \mathrm{CH_4}(s) \rightarrow \mathrm{CH_3OH}(s).
  \label{o_ins}
\end{equation}

\noindent
This type of reaction is known as an ``insertion'' since the oxygen atom is
inserted into one of the C-H bonds to form methanol.  Reaction \eqref{o_ins} is
highly endothermic, having an activation energy of $\sim4300$ K
\citep{baulch_evaluated_1992}; however, \citet{bergner_methanol_2017} recently
found that O($^1$D) and methane could efficiently react to form methanol in low
temperature ices via this mechanism. Further evidence for the importance of
solid-phase irradiation-driven insertion reactions comes from recent work by
\citet{bergantini_mechanistical_2018}, who found that such processes could lead
to ethanol and dimethyl ether formation at low-temperatures.  Thus, Class 2
reactions  may contribute to the formation of COMs, even in cold interstellar
environments.

In this study, we added Class 2 reactions for both C$^*$ and O$^*$, as listed in
Table 5 of Appendix B.  Many of these new reactions were
drawn from combustion chemistry. Since cosmic rays, such as other forms of
ionizing radiation, produce highly non-thermal species, some of the endothermic
reactions previously considered in the context of high-temperature systems
become relevant when considering irradiated low-temperature ices. 

We have also included gas-phase destruction reactions for HOCO. 
In addition to photodissociation by internal and external UV photons, the
reactions listed in Table 6 of Appendix C were added to the
\citet{ruaud_gas_2016} network, with neutral-neutral rate coefficient parameters given in terms
of $\alpha$, $\beta$, and $\gamma$  using the Arrhenius-Kooij formula

\begin{equation}
  \label{ak_formula}
  k_\mathrm{AK} = \alpha\left(\frac{T_\mathrm{gas}}{300\,\mathrm{K}}\right)^\beta \mathrm{exp}\left(-\frac{\gamma}{T_\mathrm{gas}}\right)
\end{equation}

\noindent
where $T_\mathrm{gas}$ is the kinetic temperature of the gas. 

For reactions between the polar neutral HOCO and ions, we use the Su-Chesnavich  capture theory (see \citet{woon_quantum_2009} and references therein). 
For HOCO, values of $\mu_\mathrm{D}=3.179$ D and
$\alpha_\mathrm{p}=2.739$ \AA$^3$ were utilized for the dipole and dipole polarizability,
respectively \citep{johnson_nist_2016}.

\section{Results \& Discussion} \label{sec:results}

\begin{figure}[htb!]
  \gridline{
  \fig{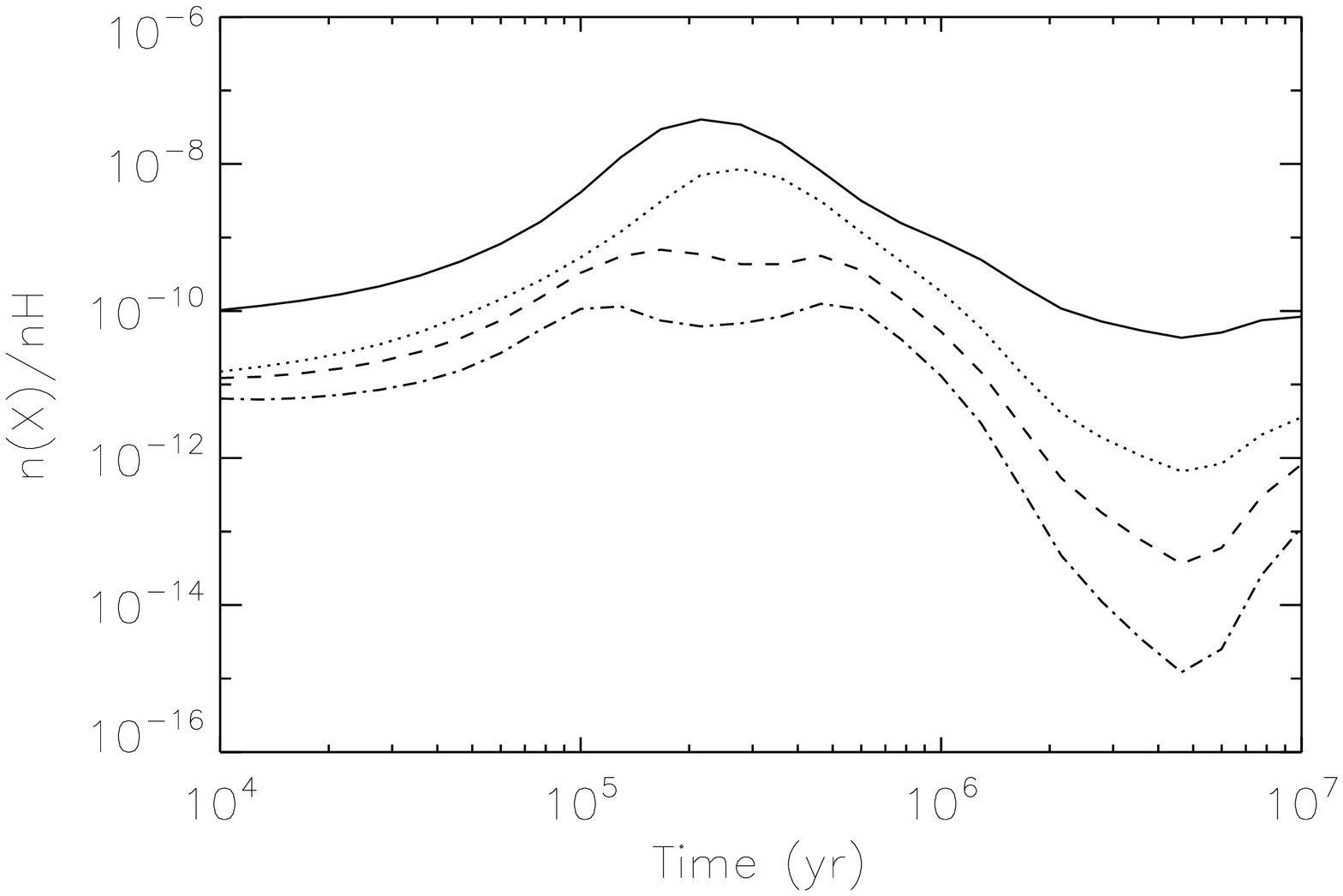}{0.48\textwidth}{(a)}
  \fig{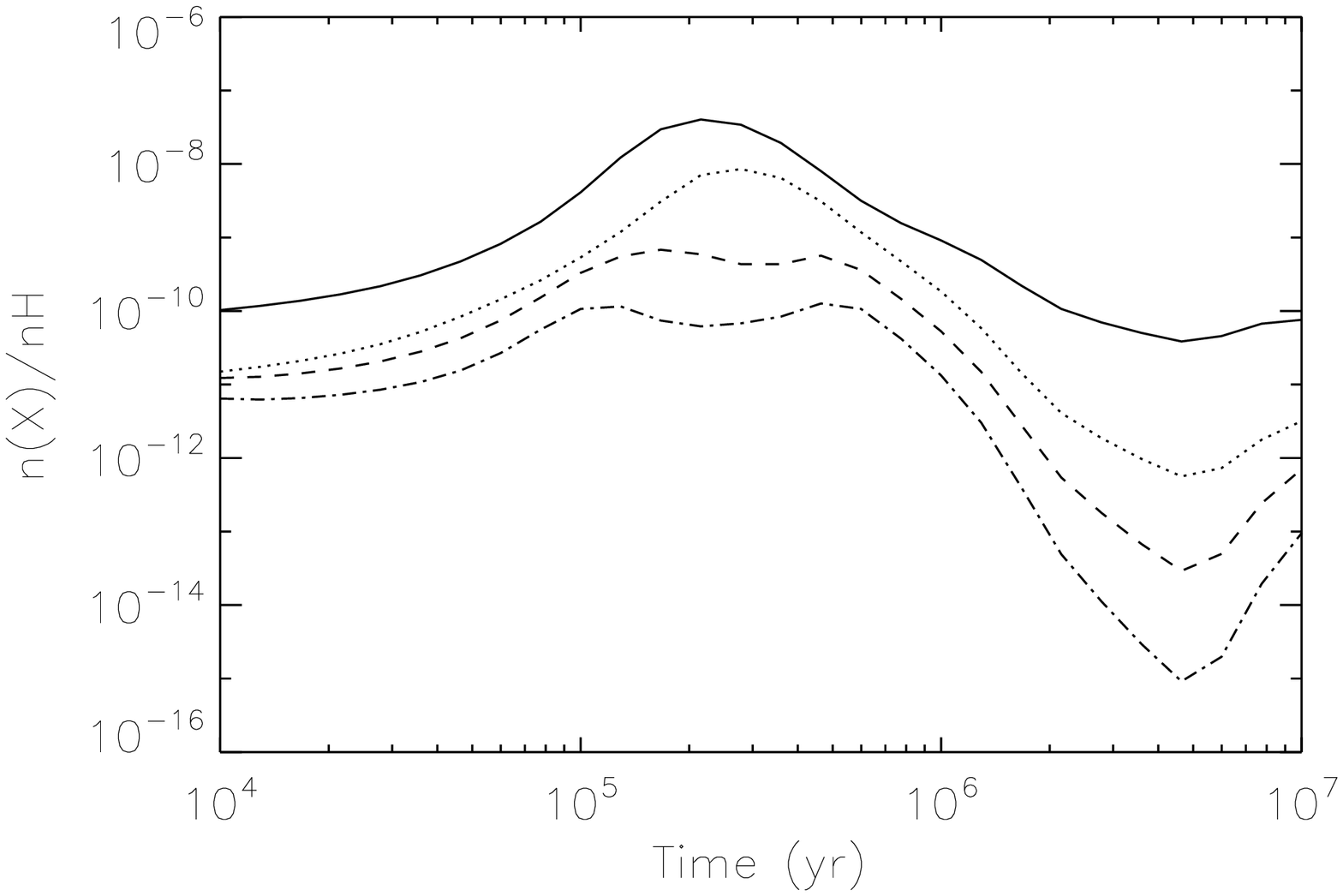}{0.48\textwidth}{(b)}
  }
  \caption{
    Gas-phase abundances in TMC-1 of HC$_3$N (solid line), HC$_5$N (dotted
    line), HC$_7$N (dashed line), and HC$_9$N (dot-dashed line), calculated
    both with (a) and without (b) the new radiation chemistry.
  }
  \label{f1}
\end{figure}

\begin{figure}[htb!]
  \gridline{
  \fig{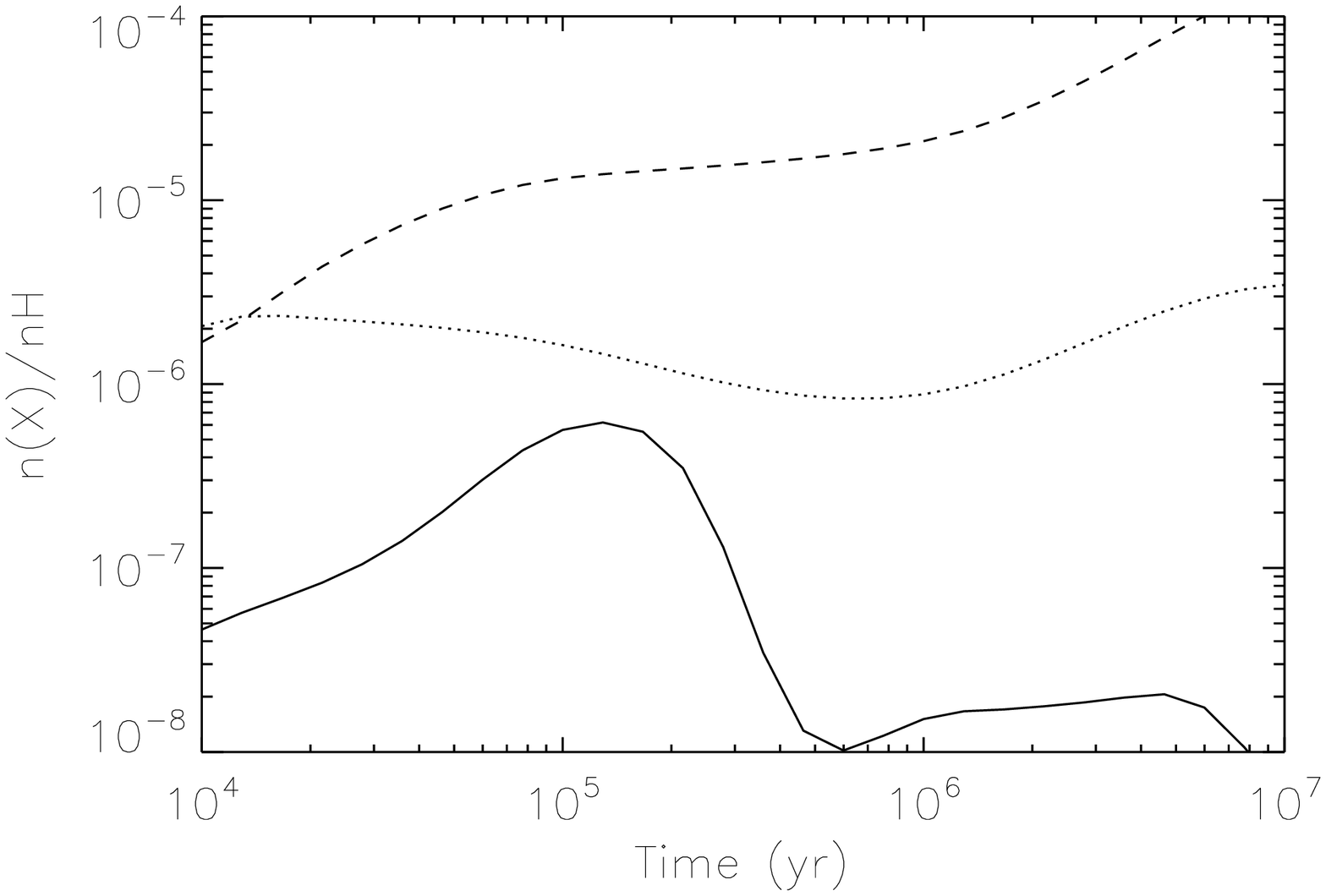}{0.48\textwidth}{(a)}
  \fig{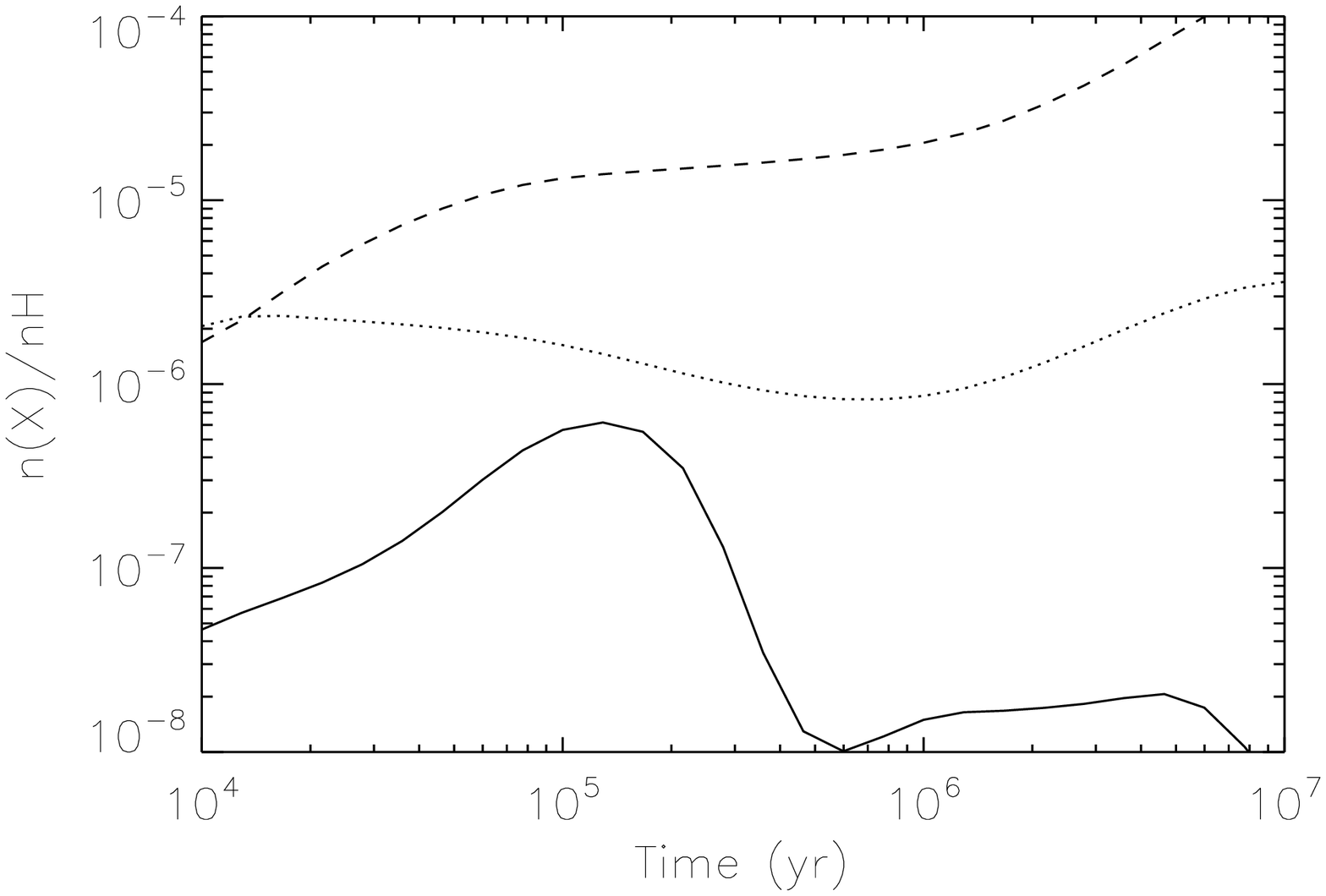}{0.48\textwidth}{(b)}
  }
  \caption{
    Abundances of H$_2$O in the gas (solid line), on grain surfaces (dotted
    line), and in the bulk (dashed line), calculated both with (a) and
    without (b) the new radiation chemistry.
  }
  \label{f2}
\end{figure}

Given the relative novelty of the radiation chemistry we have added to our chemical
network, it is natural to question what effect these new reactions will have
on the abundances of important cold core species.  To that end, in Fig.
\ref{f1} we show the calculated abundances of the cyanopolyynes in our TMC-1
models, both with and without the new reactions listed in Tables 4 and 5. 
Reassuringly, one can see that there are no significant differences between
cyanopolyyne abundances in the two sets of results - a key test since
modern astrochemical models are typically able to reproduce the observed
abundances of these species quite well \citep{mcguire_detection_2017} 

Since cyanopolyynes are formed in the gas-phase
\citep{loomis_non-detection_2016}, and all of the radiolysis processes
considered in this work are assumed to take place in or on the surface of
dust-grain ice mantles, a better confirmation of the new chemistry may be
to examine the abundance of the primary ice-mantle constituent,
namely, water.  Therefore, in Fig. \ref{f2} we show the abundance of water in
the gas-phase, ice-surface, and ice-bulk in our TMC-1 models both with and
without radiation chemistry. Again, we find that the differences between the
two are negligible. Thus, the addition of the novel reactions does not lead to
unphysical predictions for common species (e.g. water), nor does it obviously
degrade our ability to reproduce the abundances of commonly observed molecules
such as the cyanopolyynes. 

However, we have found that the addition of cosmic ray-driven reactions does indeed have
a significant effect on the abundances of a number of astrochemically interesting species in our model.
In the remainder of this section, we will describe how the inclusion of radiation chemistry
affects HOCO, NO$_2$, HC$_2$O, and HCOOCH$_3$, which showed the most
pronounced enhancements in gas-phase abundance. 

\FloatBarrier

% HOCO
\subsection{HOCO}

\begin{figure}[htb!]
  \gridline{
  \fig{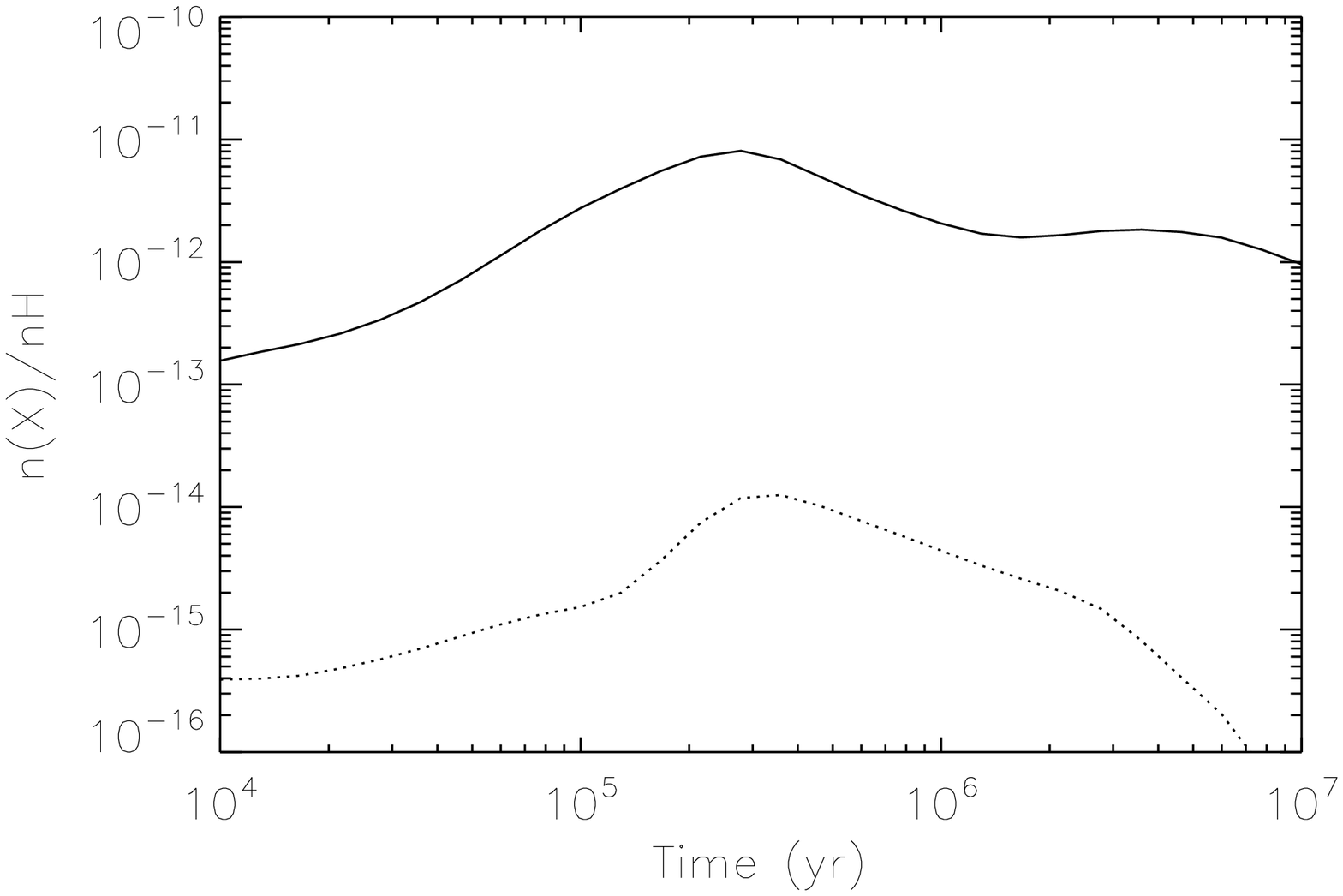}{0.5\textwidth}{(a)}
  \fig{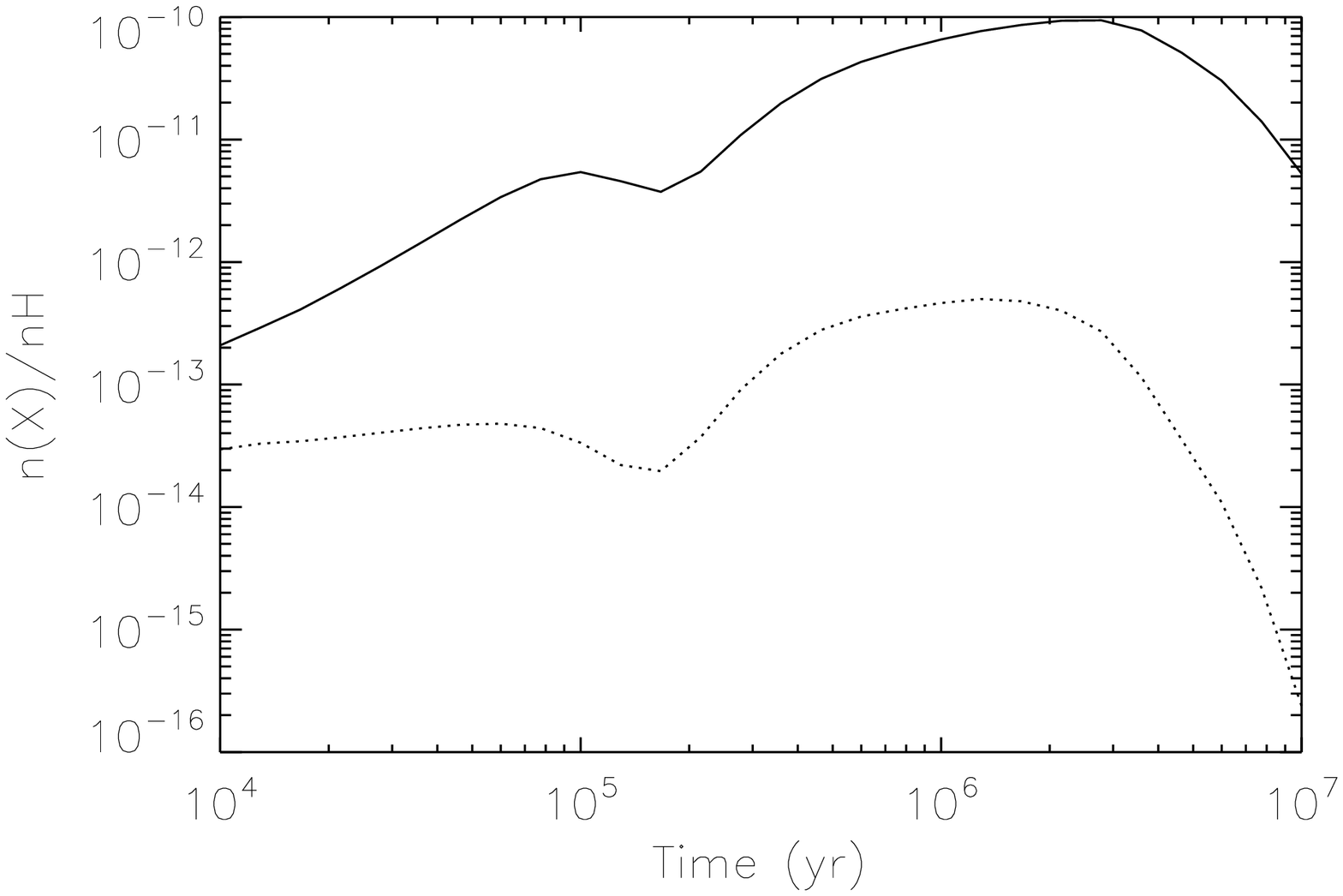}{0.5\textwidth}{(b)}
  }
  \gridline{
  \fig{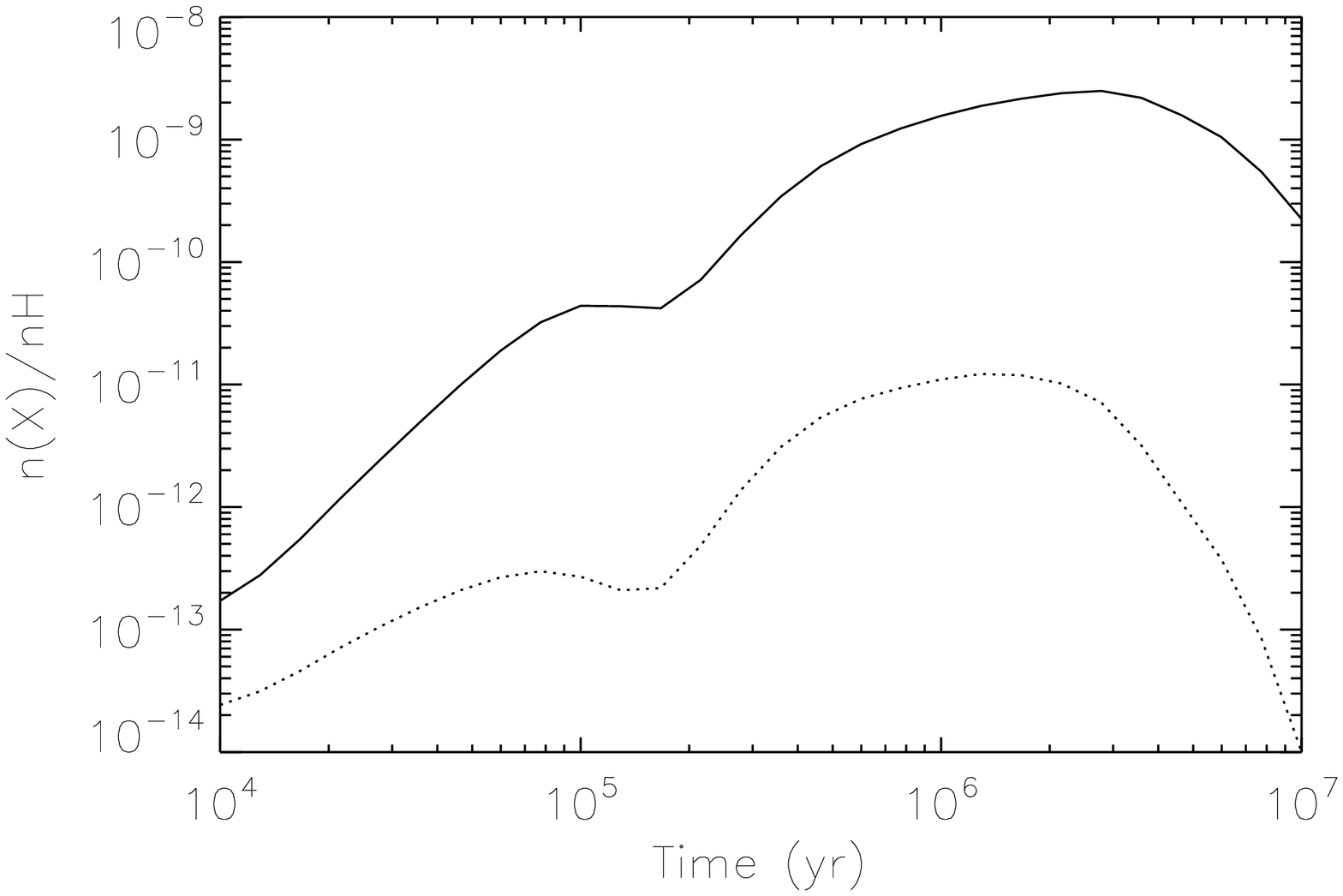}{0.5\textwidth}{(c)}
  }
  \caption{
  Simulated TMC-1 abundances of HOCO in the gas (a), on the grain/ice surface
  (b), and in the ice bulk (c), calculated both with (solid line) and without
  (dotted line) radiation chemistry.
  }
  \label{f3}
\end{figure}

As shown in Fig. \ref{f3}, abundances of HOCO are increased in simulations
including radiation chemistry in the three phases of the model:  gas, ice
surface, and ice bulk. This increase is due primarily to the following surface
reaction

\begin{equation}
 \mathrm{OH^*}(s)+ \mathrm{CO}(s) \rightarrow \mathrm{HOCO}(g)
\end{equation}

\noindent
where the HOCO product undergoes chemical desorption
\citep{garrod_non-thermal_2007}. Here, surface abundances of CO are primarily
the result of the adsorption from the gas-phase, and OH$^*$ is primarily formed
via the Type I radiolysis of water

\begin{equation}
 \mathrm{H_2O}(s) \leadsto \mathrm{OH^*}(s) + \mathrm{H^*}(s).
 \label{watertooh}
\end{equation}

The fact that HOCO is significantly enhanced in our TMC-1 simulations is
notable because this species is more commonly encountered in high-temperature
combustion chemistry \citep{smith_rate_1973,mccarthy_isotopic_2016}; however,
in \citet{milligan_infrared_1971} - perhaps the first work to identify HOCO -
this species was detected in a mixed H$_2$O:CO ice after irradiation by VUV
photons, underscoring the similarity between the products of both combustion
and radiation (or high-energy photo-) chemistry. Thus, the detection of species
like HOCO in a cold interstellar region would be a strong indication of cosmic
ray-induced radiation chemistry at work. 

As shown in Fig. \ref{f3}a, the peak gas-phase fractional abundance of HOCO is
$\sim10^{-11}$. Assuming a hydrogen column density for TMC-1 of
$N(\mathrm{H_2})\approx 10^{22}$ cm$^{-2}$ \citep{gratier_new_2016} results in
a predicted HOCO column of $N(\mathrm{HOCO}) \approx 10^{11}$ cm$^{-2}$.
Since
HOCO has a dipole of $\sim 3$ Debye \citep{johnson_nist_2016}, these model
results imply that this species is potentially observable in TMC-1. 

% NO2
\subsection{NO$_2$}

\begin{figure}[htb!]
  \gridline{
  \fig{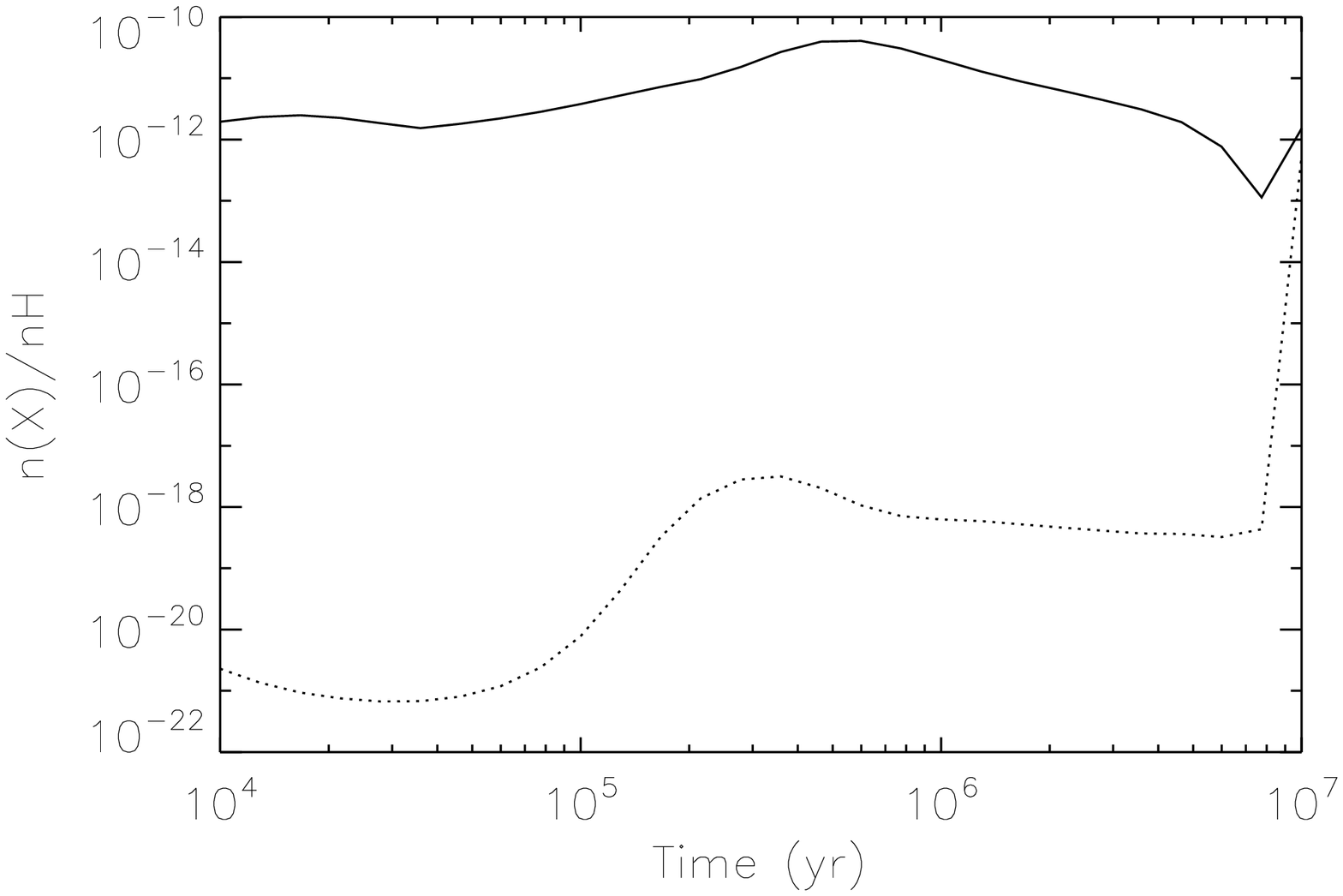}{0.5\textwidth}{(a)}
  \fig{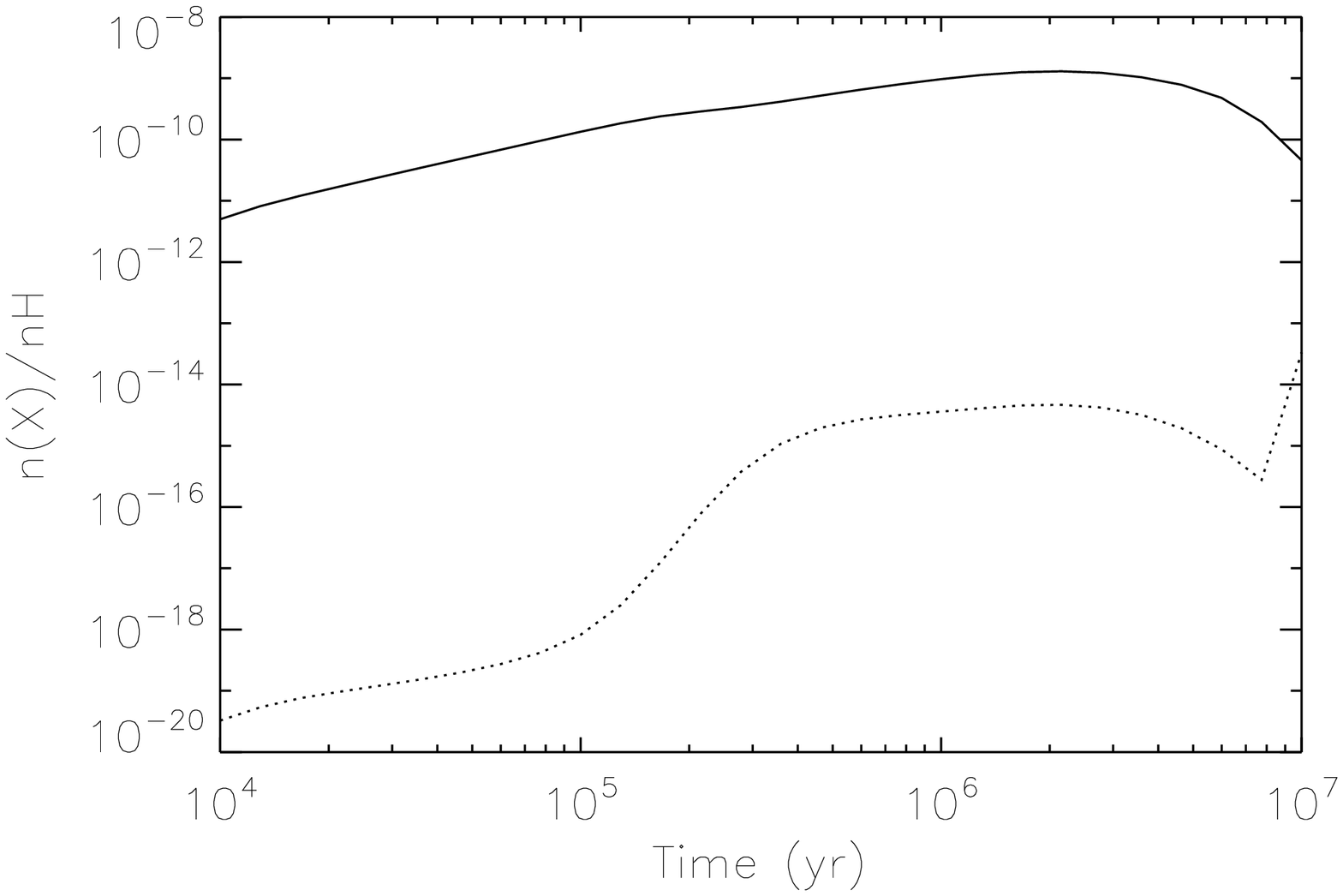}{0.5\textwidth}{(b)}
  }
  \gridline{
  \fig{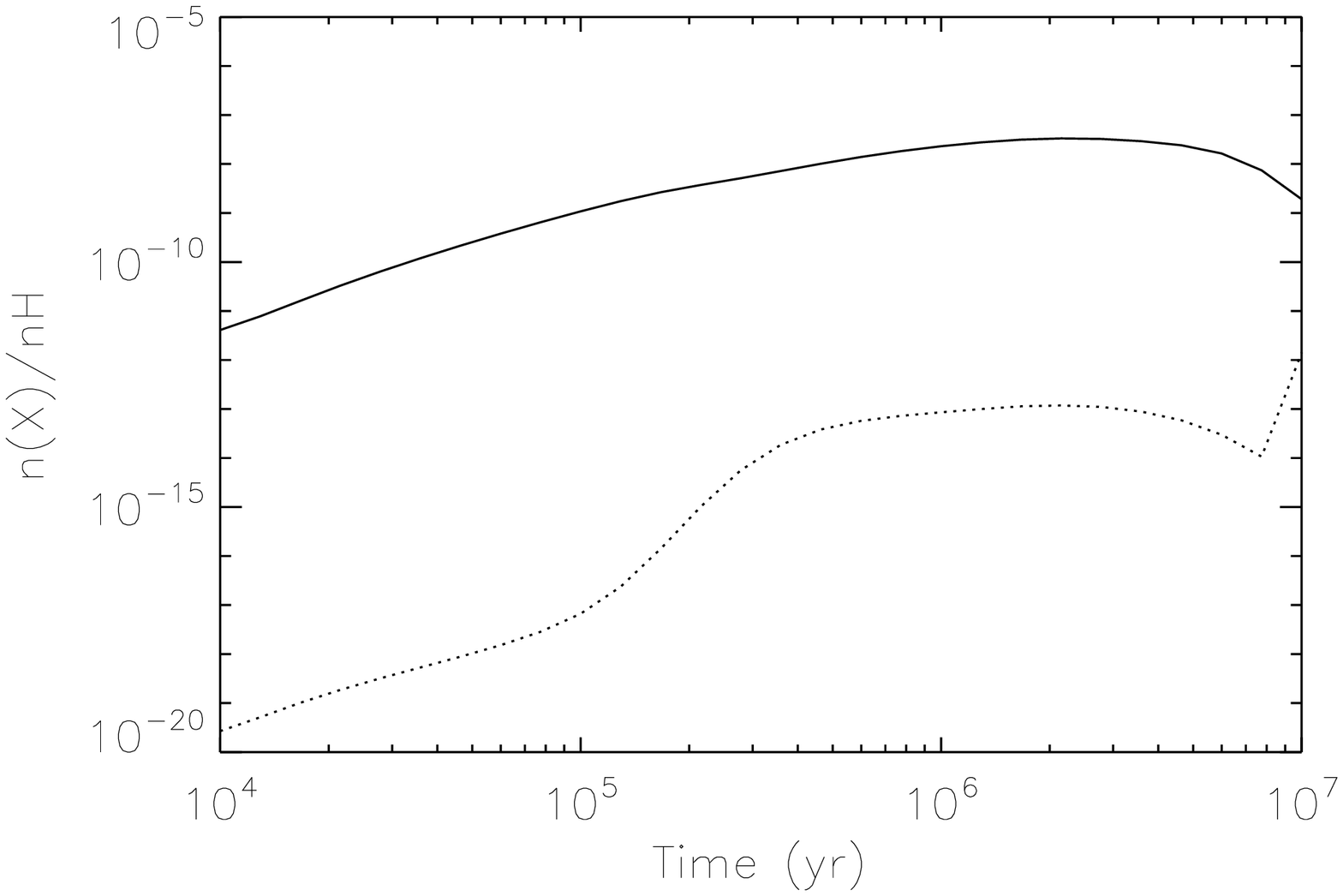}{0.5\textwidth}{(c)}
  }
  \caption{
  Simulated TMC-1 abundances of NO$_2$ in the gas (a), on the grain/ice surface
  (b), and in the ice bulk (c), calculated both with (solid line) and without
  (dotted line) radiation chemistry.
  }
  \label{f4}
\end{figure}

NO$_2$ is another species the abundance of which is enhanced in simulations that
include radiation chemistry. As shown in Fig. \ref{f4}, as for the case of HOCO, NO$_2$
abundances are increased in all three model phases, although the connection
between these enhancements and radiation chemistry is slightly more complex
than in the case of HOCO. 

At early times ($<10^4$ yr), the dominant formation route for gas-phase NO$_2$
is the reaction

\begin{equation}
  \mathrm{NO}(g) + \mathrm{O_2H}(g) \rightarrow \mathrm{NO_2}(g) + \mathrm{OH}(g).
\end{equation}

\noindent
Here, gas-phase O$_2$H abundances are enhanced via

\begin{equation}
  \mathrm{O}(s) + \mathrm{OH^*}(s) \rightarrow \mathrm{O_2H}(g).
\end{equation}

At later times in the TMC-1 simulations, the dominant formation routes for
NO$_2$ are 

\begin{equation}
  \mathrm{O}(s) + \mathrm{NO^*}(s) \rightarrow \mathrm{NO_2}(g)
\end{equation}

\begin{equation}
  \mathrm{O^*}(s) + \mathrm{NO}(s) \rightarrow \mathrm{NO_2}(g).
\end{equation}

\noindent
At all simulation times, surface NO$^*$ is formed mainly via the Type III
excitation of NO:

\begin{equation}
  \mathrm{NO}(s) \leadsto \mathrm{NO^*}(s)
\end{equation}

\noindent
while O$^*$ is formed from the Type I radiolysis of water, CO, and CO$_2$:

\begin{equation}
\mathrm{H_2O}(s) \leadsto \mathrm{O^*}(s) + \mathrm{H_2^*}(s),
\end{equation}

\begin{equation}
  \mathrm{CO}(s) \leadsto \mathrm{C^*}(s) + \mathrm{O^*}(s),
\end{equation}

\begin{equation}
  \mathrm{CO_2}(s) \leadsto \mathrm{CO^*}(s) + \mathrm{O^*}(s).
\end{equation}

As shown in Fig. \ref{f4}a, the peak gas-phase relative abundance of NO$_2$ in
our TMC-1 model is $\sim4\times10^{-11}$, corresponding to a column density of
$\sim 4\times10^{11}$ cm$^{-2}$. Though this is slightly higher than the
predicted abundance of HOCO, observations of NO$_2$ are challenging due to its
small permanent dipole of $<1$ Debye \citep{johnson_nist_2016}.

% HC2O
\subsection{HC$_2$O}

\begin{figure}[htb!]
  \gridline{
  \fig{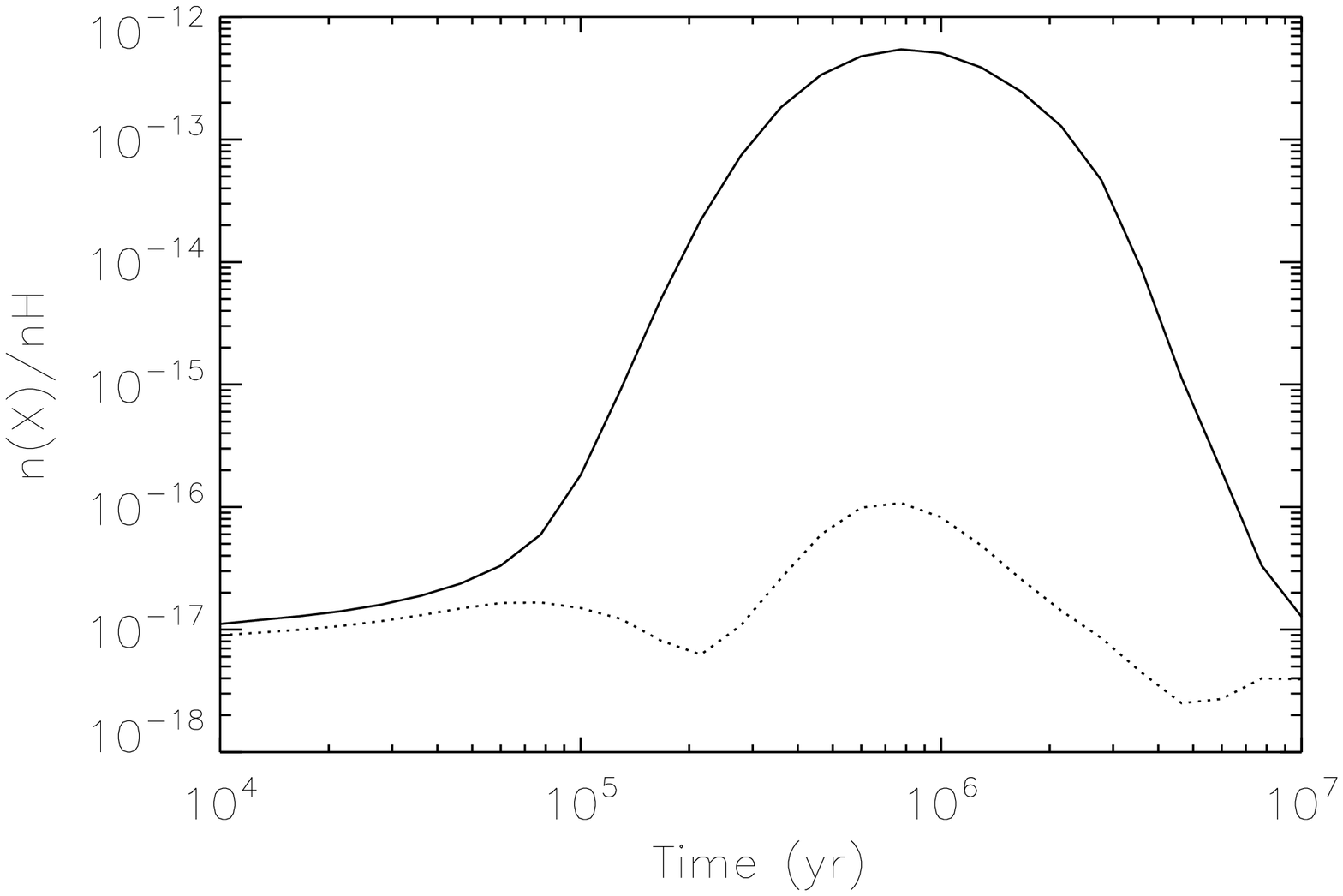}{0.5\textwidth}{(a)}
  \fig{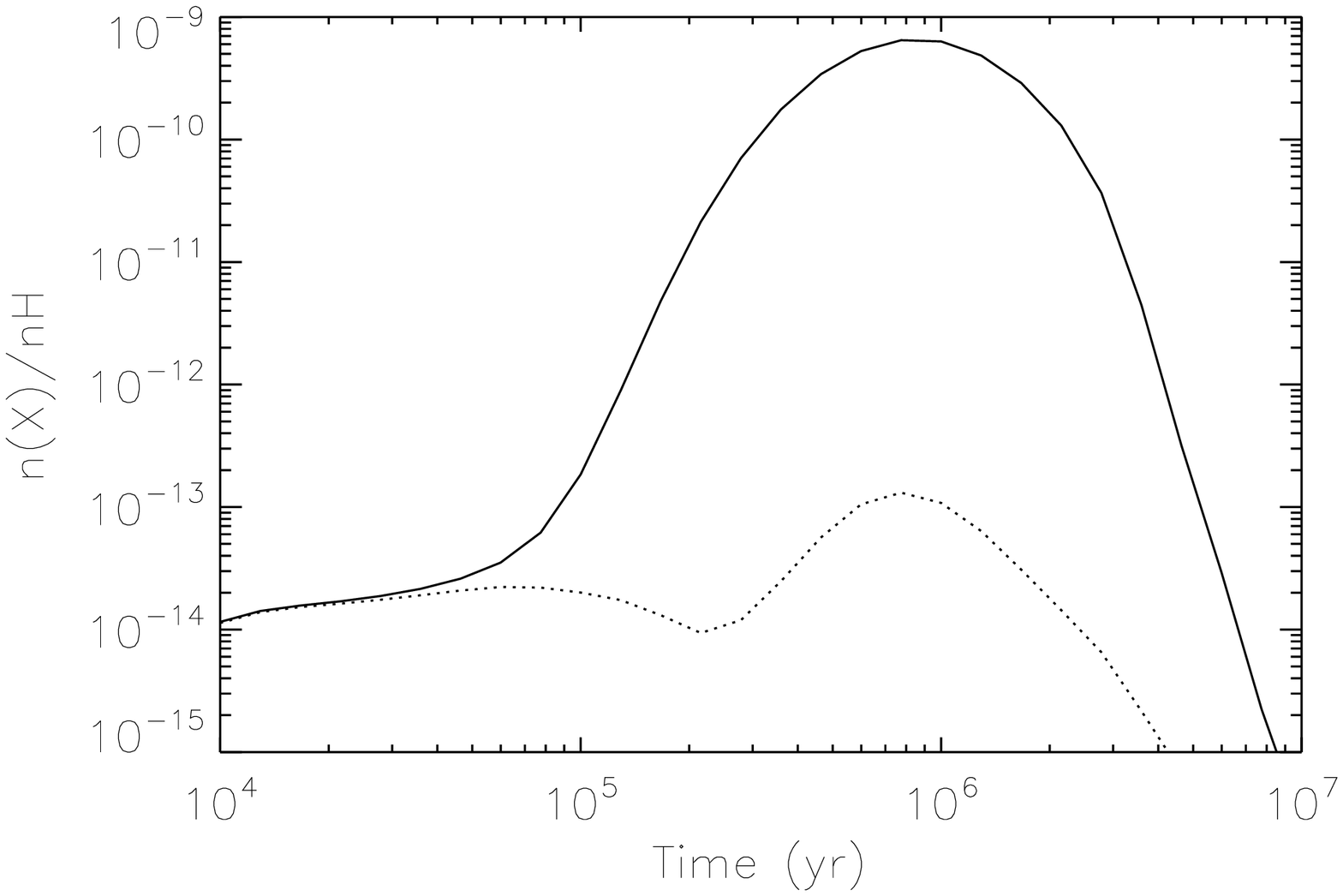}{0.5\textwidth}{(b)}
  }
  \gridline{
  \fig{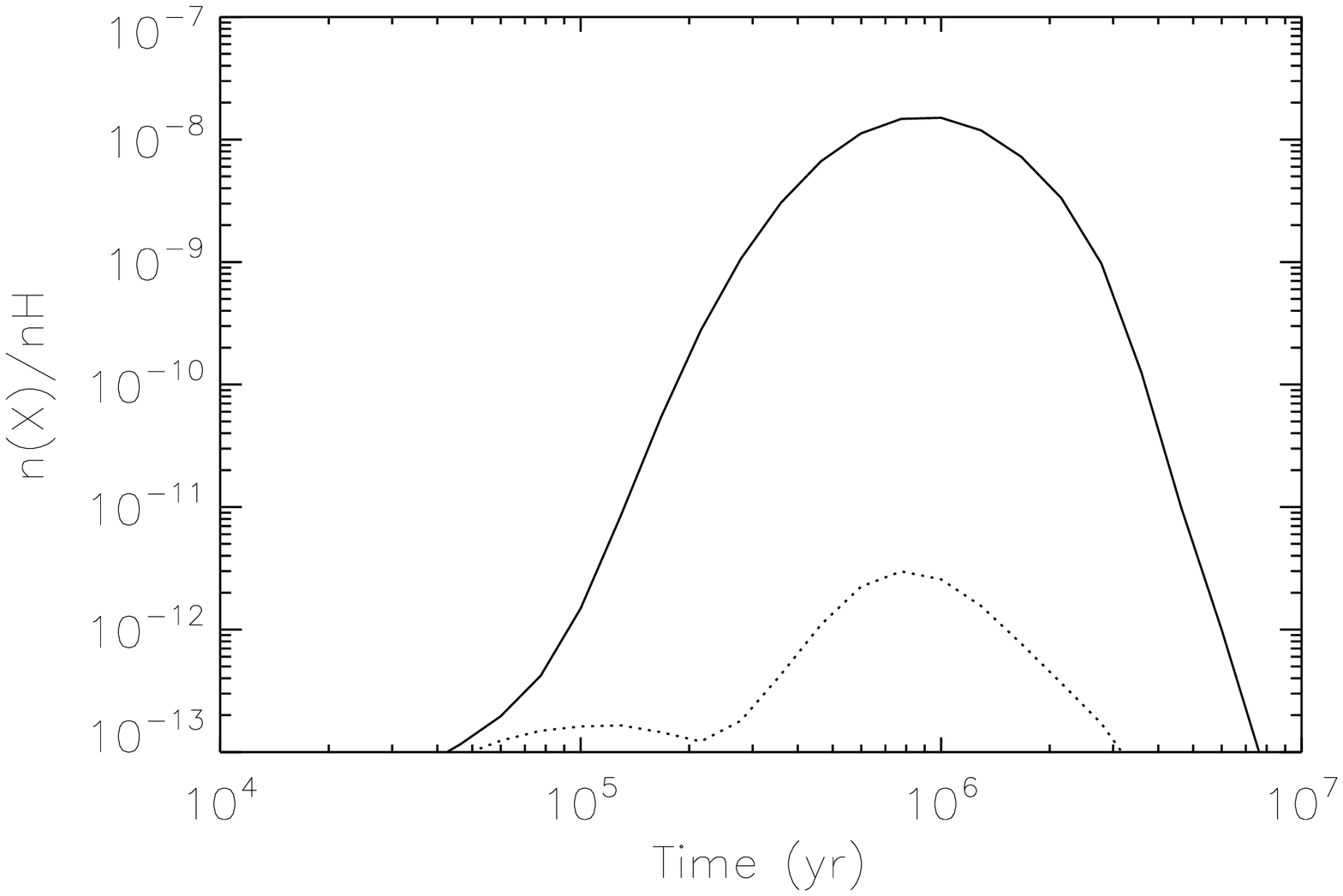}{0.5\textwidth}{(c)}
  }
  \caption{
  Simulated TMC-1 abundances of HC$_2$O in the gas (a), on the grain/ice
  surface (b), and in the ice bulk (c), calculated both with (solid line) and
  without (dotted line) radiation chemistry.
  }
  \label{f5}
\end{figure}

The ketenyl radical, HC$_2$O, was first observed in the cold ($T_\mathrm{kin}
\approx 15$ K) starless cores Lupus-1A and L483 by
\citet{agundez_discovery_2015}, who derived a column density of $\sim 5\times
10^{11}$ cm$^{-2}$ for both sources. Chemical simulations
were run assuming HC$_2$O formation via the reaction of OH and C$_2$H.
It was noted that such simulations underproduce the
ketenyl radical by about six orders of magnitude, leading the authors to posit the
existence of ``a powerful formation mechanism'' to counterbalance HC$_2$O
destruction pathways. 

As shown in Fig. \ref{f5}, the inclusion of radiation chemistry in our TMC-1
simulations results in significant enhancements of HC$_2$O - roughly four
orders of magnitude for the gas, ice surface, and ice bulk. At early simulation
times ($<10^3$ yr), the dominant formation route for gas-phase ketenyl radical
is

\begin{equation}
  \mathrm{OH^*}(s) + \mathrm{CCH}(s) \rightarrow \mathrm{HC_2O}(g) + \mathrm{H}(g).
\end{equation}

\noindent
At all later simulation times ($>10^3$ yr), HC$_2$O is mainly formed via

\begin{equation}
  \mathrm{H}(s) + \mathrm{CCO}(s) \rightarrow \mathrm{HC_2O}(g).
\end{equation}

\noindent
In both TMC-1 simulations with and without radiation chemistry, there is little
difference in the CCH abundance at all times and for all phases of the model;
however, the ice surface and bulk abundances of CCO are enhanced via the
reaction

\begin{equation}
  \mathrm{C^*}(s) + \mathrm{CO}(s) \rightarrow \mathrm{CCO}(s)
\end{equation}

\noindent
where the suprathermal carbon atoms are formed mainly via the radiolysis
of CO. 

Though our simulations still underproduce gas-phase HC$_2$O compared with observed values of \citet{agundez_discovery_2015}, the significant enhancements seen in models run with radiation chemistry  suggest that perhaps radiation chemistry is  their speculated powerful formation mechanism.  Since we have not included any non-thermal desorption mechanisms caused by the direct cosmic ray bombardment of dust grains, such as sputtering, it may be that the impact of radiation chemistry on gas-phase abundances
is greater than what is implied by our results here. 

% METHYL FORMATE
\subsection{HCOOCH$_3$}

\begin{figure}[htb!]
  \gridline{
  \fig{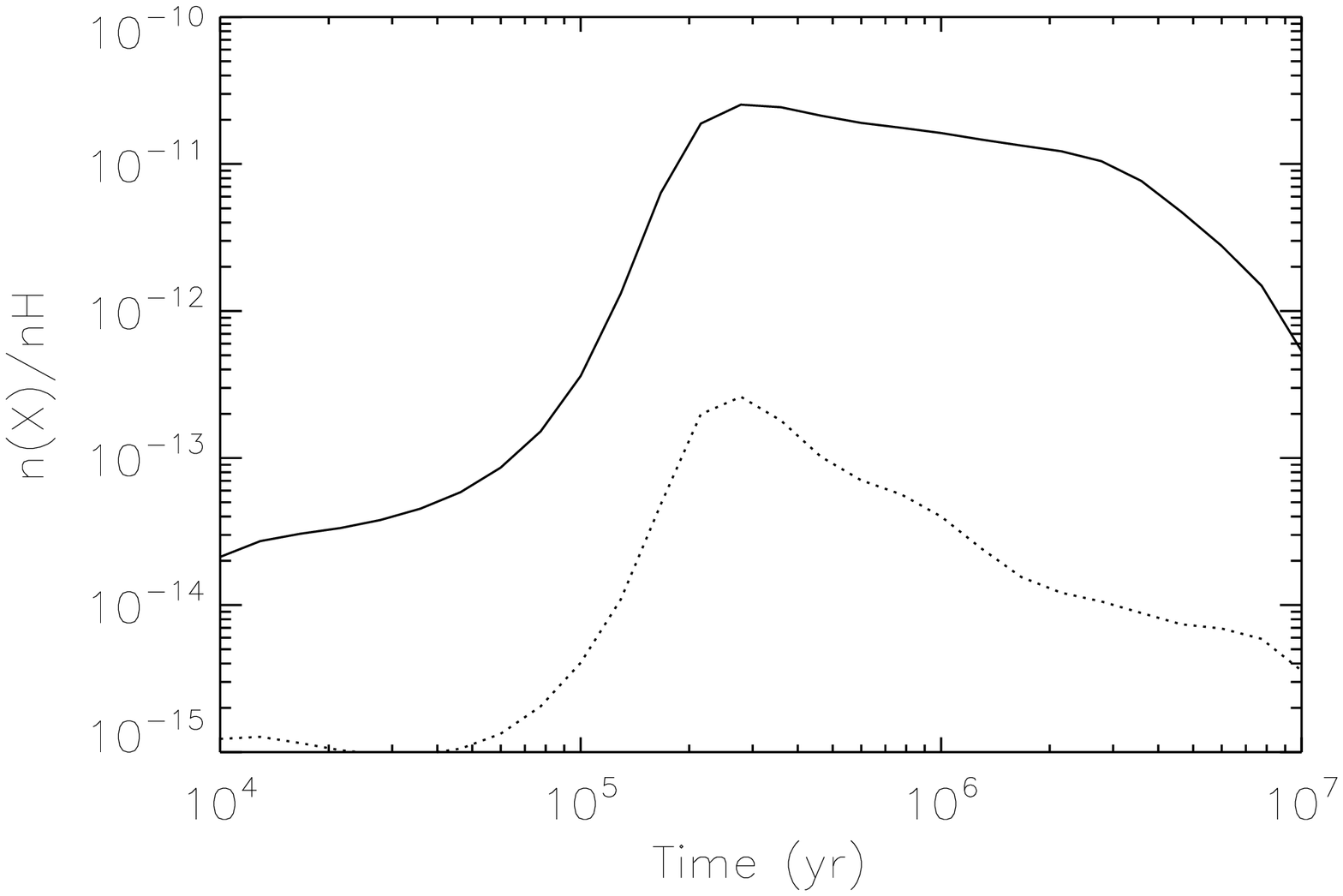}{0.5\textwidth}{(a)}
  \fig{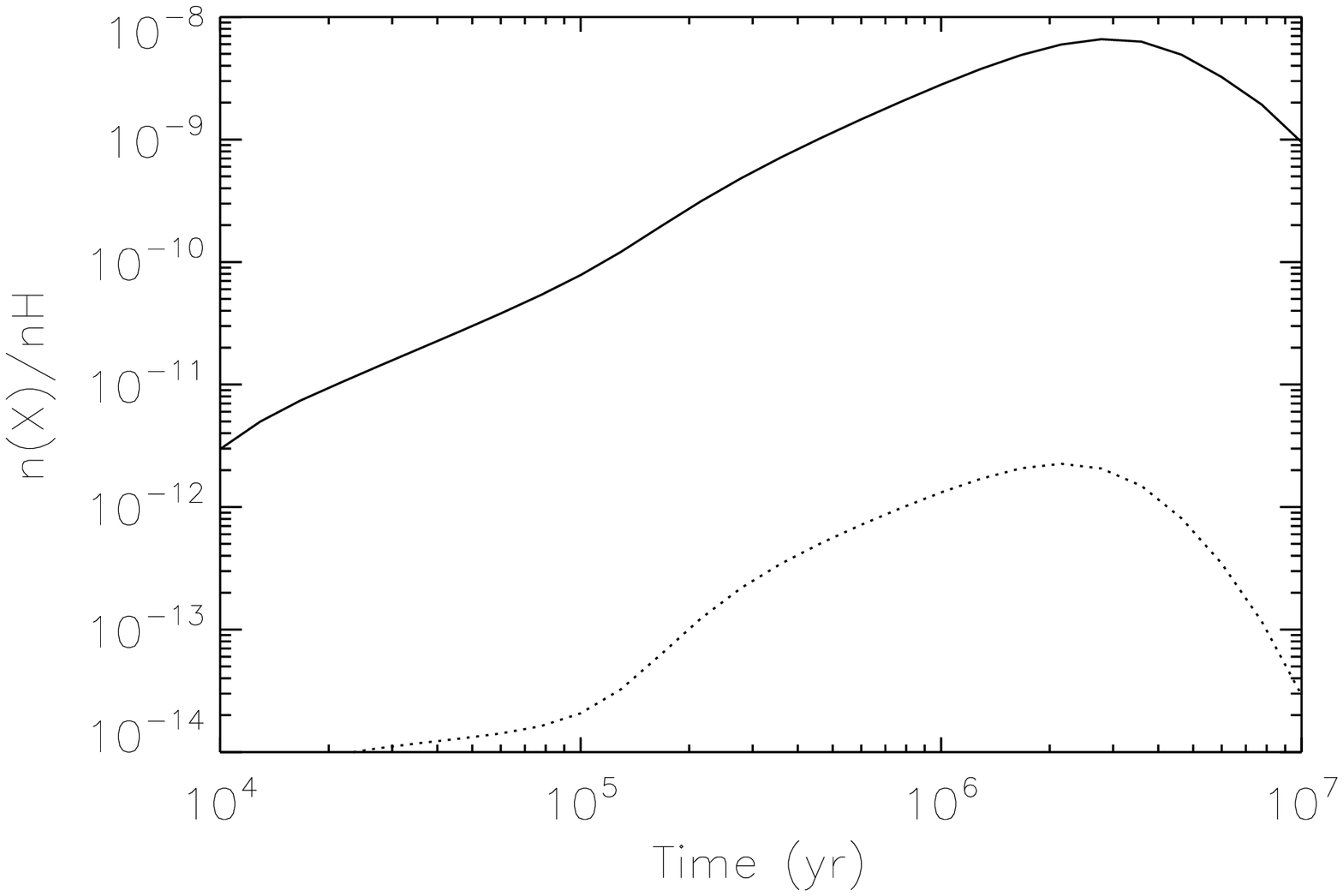}{0.5\textwidth}{(b)}
  }
  \gridline{
  \fig{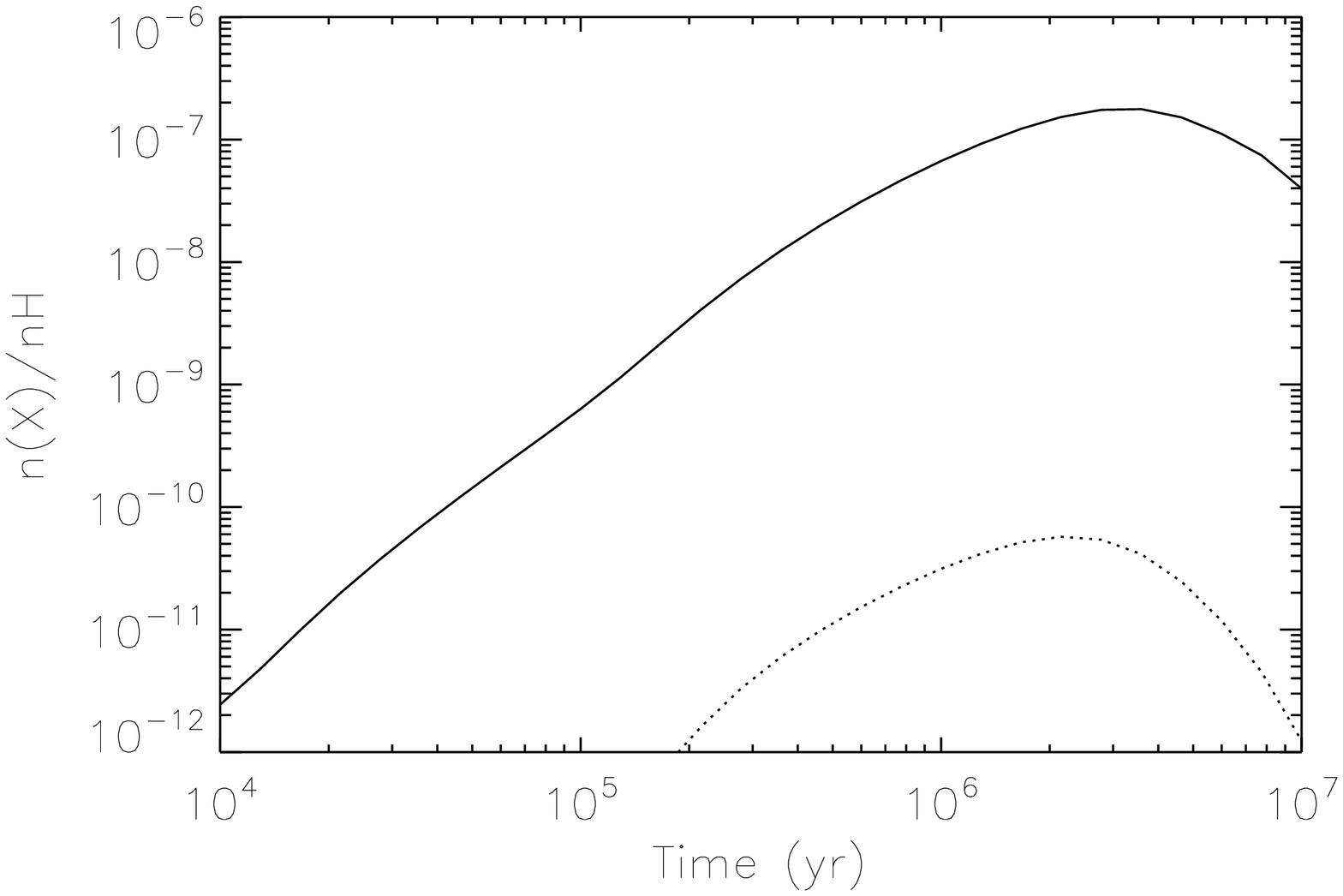}{0.5\textwidth}{(c)}
  }
  \caption{
  Simulated TMC-1 abundances of HCOOCH$_3$ in the gas (a), on the grain/ice
  surface (b), and in the ice bulk (c), calculated both with (solid line)
  and without (dotted line) radiation chemistry.
  }
  \label{f6}
\end{figure}

As with the other species highlighted thus far, the abundance of methyl formate
(HCOOCH$_3$) is enhanced in all three phases of the model. In simulations
including radiation chemistry, the main production pathways for gas-phase
methyl formate are

\begin{equation}
  \mathrm{HCO^*}(s) + \mathrm{CH_3O}(s) \rightarrow \mathrm{HCOOCH_3}(g)
\end{equation}
and
\begin{equation}
  \mathrm{HCO}(s) + \mathrm{CH_3O^*}(s) \rightarrow \mathrm{HCOOCH_3}(g).
\end{equation}

\noindent
Here, the suprathermal HCO is produced mainly via the Type I radiolysis of 
formaldehyde:

\begin{equation}
  \mathrm{H_2CO}(s) \leadsto \mathrm{H^*}(s) + \mathrm{HCO^*}(s)
\end{equation}

\noindent
and the methoxy radical is produced from the Type I decomposition of methanol:

\begin{equation}
  \mathrm{CH_3OH}(s) \leadsto \mathrm{H^*}(s) + \mathrm{CH_3O^*}(s).
\end{equation}

Methyl formate has been a focus of several 
recent studies which likewise examined
its formation in cold cores \citep{balucani_formation_2015,chang_unified_2016,vasyunin_reactive_2013}.
In \citet{balucani_formation_2015} gas-phase production via 

\begin{equation}
  \mathrm{O} + \mathrm{CH_3OCH_2} \rightarrow \mathrm{HCOOCH_3} + \mathrm{H}
  \label{mf_balucani}
\end{equation}

\noindent
was considered. As shown in Fig. \ref{f6}, our models predict a peak gas-phase relative abundance of $\sim 3\times10^{-11}$ for methyl formate. Our peak value here is $\sim500$\% larger than than the $\sim5\times10^{-12}$ obtained by Balucani and coworkers in models 
 where they used the standard chemical desorption fraction of 1\%,  the efficiency we assume throughout this work.  Similarly,
\citet{chang_unified_2016} achieved somewhat higher gas-phase abundances of
methyl formate in a number of their cold core simulations; however, they found
that such results required both an enhanced chemical desorption fraction of
10\% and the addition of a novel ``chain reaction mechanism'' that is not
easily implemented in the macroscopic model we have utilized. 

Though the number of grain-surface formation routes for COMs like methyl
formate are limited in our network - compared with those used in hot core
simulations \citep{garrod_exploring_2017} - these results suggest radiation-chemical reactions may be able to drive the formation of COMs even
under cold core conditions. As shown, the production of these complex species
is possible because of the suprathermal reactants which form as a result of the
radiolytic dissociation of molecules in dust grain ice-mantles. 

\subsection{CH$_3$CH$_2$OH}

\begin{figure}[htb!]
  \centering
  \label{f7}
  \includegraphics[width=0.8\textwidth]{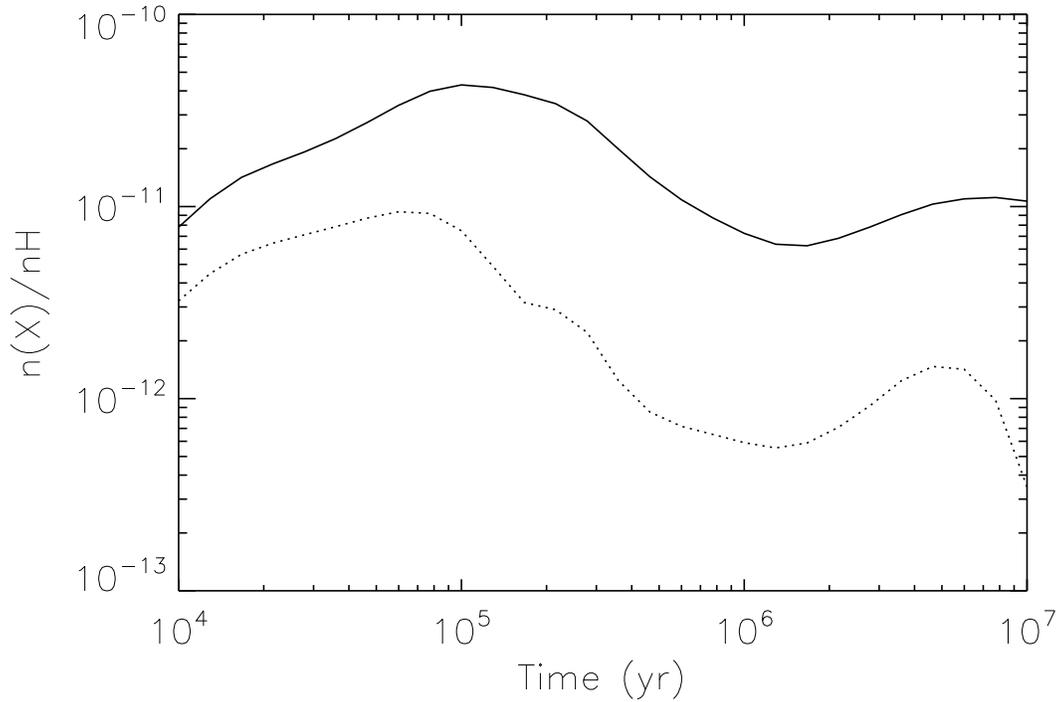}
  \caption{
    Simulated TMC-1 abundance of gas-phase ethanol (CH$_3$CH$_2$OH), 
    calculated both with (solid line) and without (dotted line) radiation
    chemistry.
  }
\end{figure}

Unlike the other species highlighted thus far, surface and bulk abundances of
the COM, ethanol, were not significantly enhanced in our simulations including
radiation chemistry. However, as shown in Fig. 7, the gas-phase abundance
is enhanced by ca. an order of magnitude by the Class 2 insertion reaction

\begin{equation}
  \mathrm{CH_2^*}(s) + \mathrm{CH_3OH}(s) \rightarrow \mathrm{CH_3CH_2OH}(g).
  \label{ralf_alcohol}
\end{equation}

\noindent
This insertion reaction, which was recently studied experimentally by
\citet{bergantini_mechanistical_2018}, was shown to efficiently form both
ethanol - as well as dimethyl ether - in low temperature ices. In that work,
Bergantini and coworkers found that CH$_2^*$ was formed from the radiolytic
decomposition of methane:

\begin{equation}
  \mathrm{CH_4}(s) \leadsto \mathrm{CH_2^*}(s) + \mathrm{H_2}(s).
\end{equation}
  
\noindent
This process, which we have included in our network, is the dominant formation
route of CH$_2^*$ at all model times. The results shown in Fig. 7
highlight the effect that Class 2 reactions such as insertions can have on the
production of COMs in cold sources. Again, we note that since chemical
desorption at the standard 1\% efficiency is the dominant non-thermal
desorption mechanism in our model, the influence of reaction
\eqref{ralf_alcohol}, and similar surface reactions, is likely underestimated
here. 

\subsection{Results Using Enhanced Ionization Rates}

\begin{figure}[htb!]
  \gridline{
  \fig{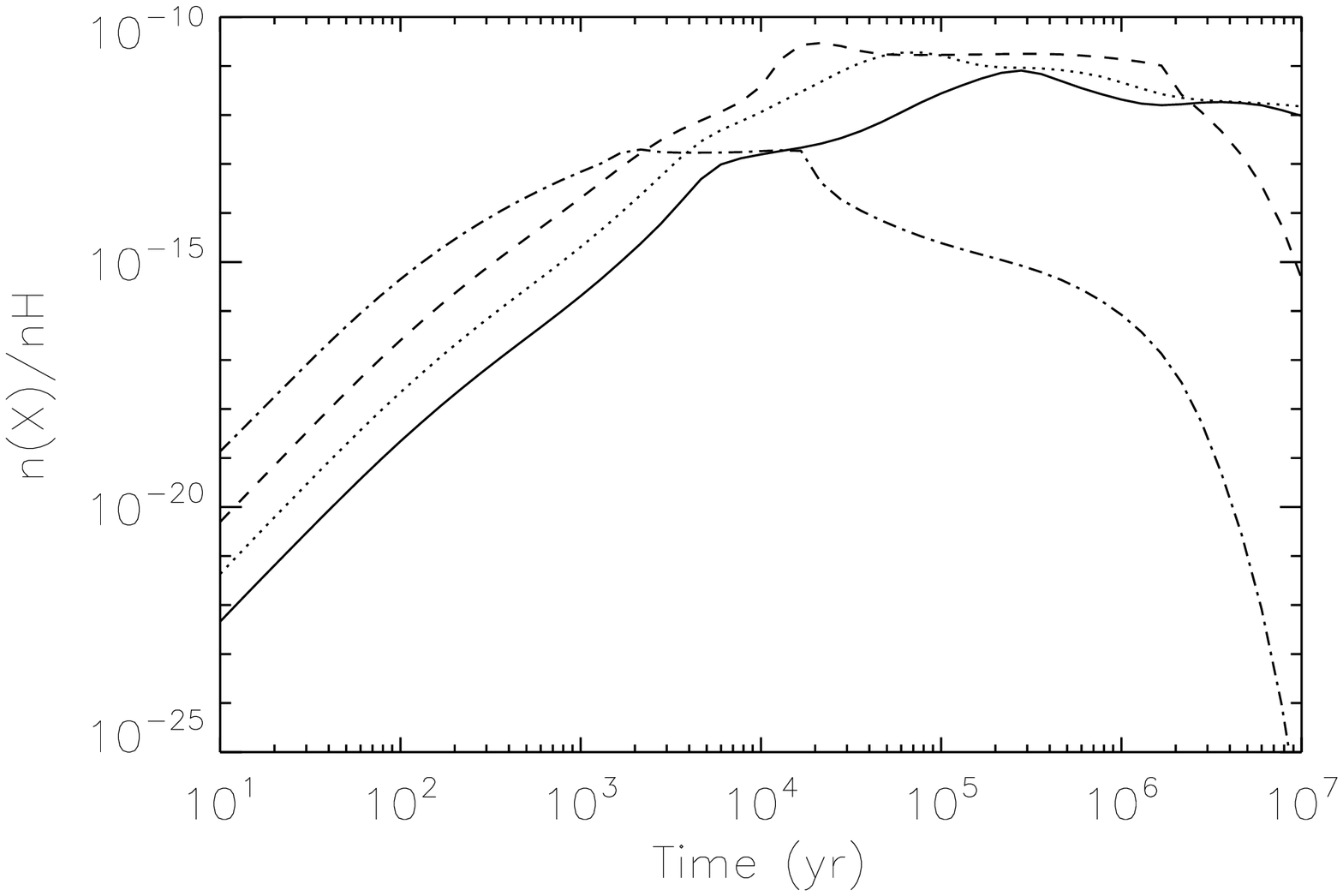}{0.5\textwidth}{(a)}
  \fig{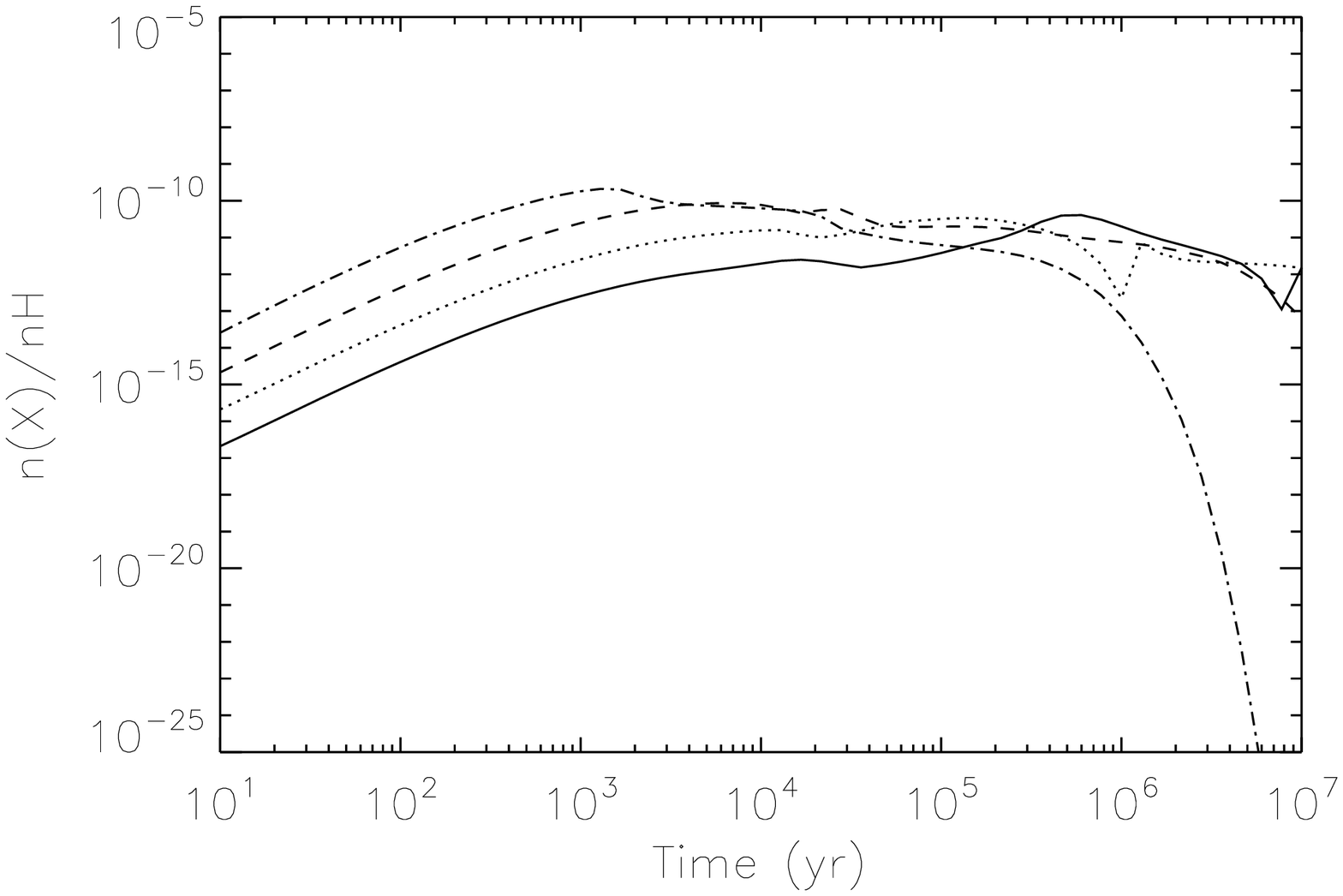}{0.5\textwidth}{(b)}
  }
  \gridline{
  \fig{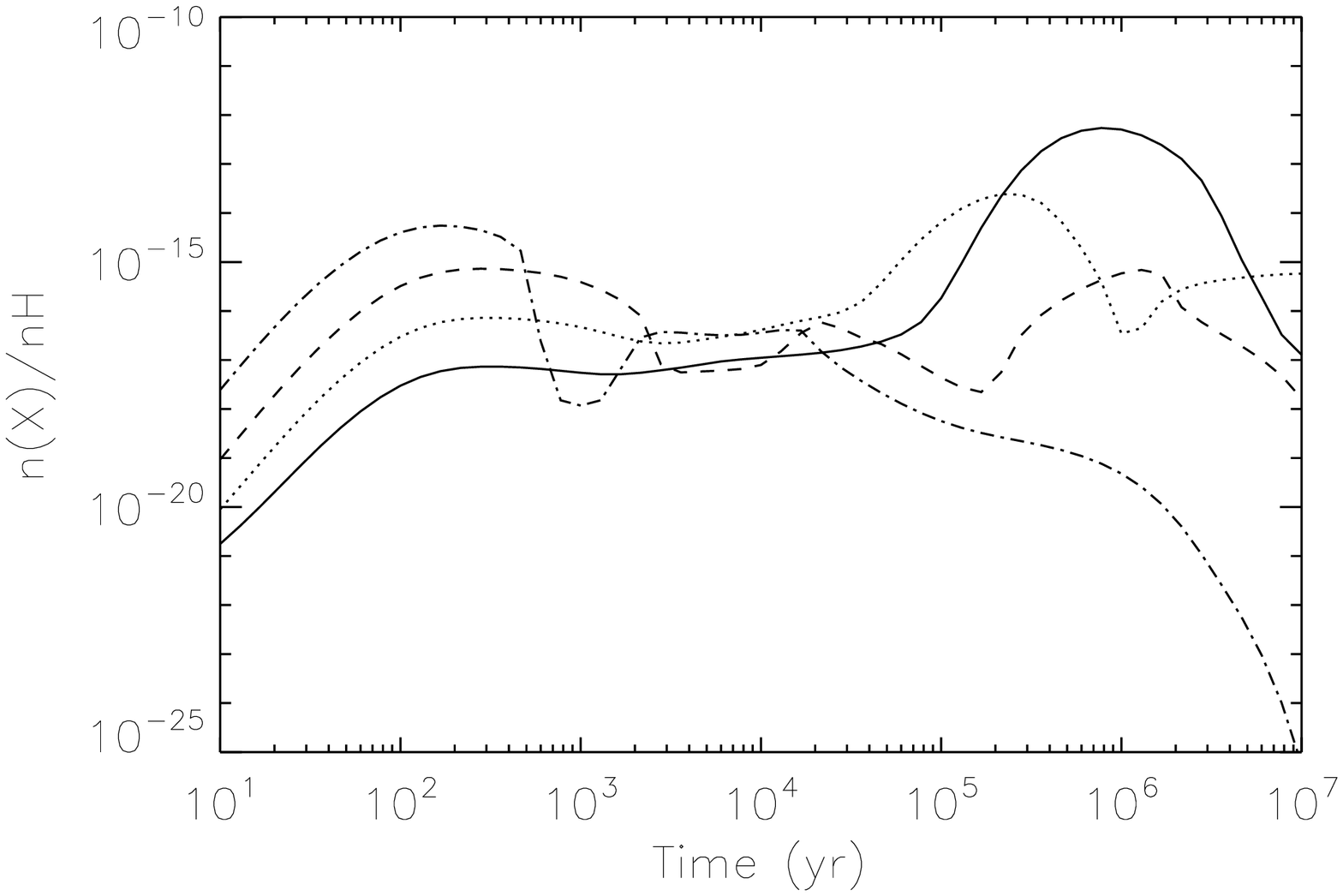}{0.5\textwidth}{(c)}
  \fig{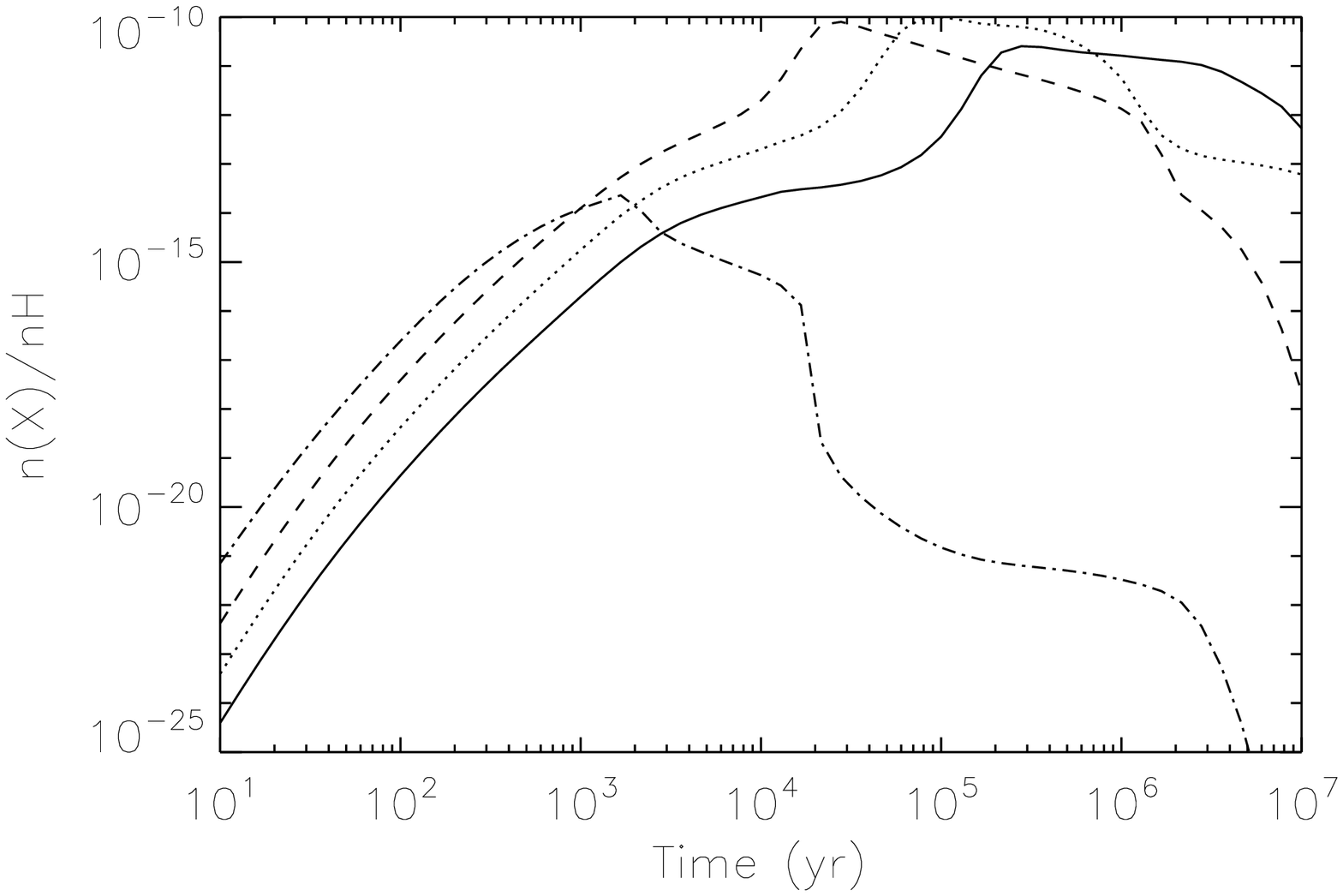}{0.5\textwidth}{(d)}
  }
  \caption{
    Calculated gas-phase abundances of HOCO (a) NO$_2$ (b) HC$_2$O (c), and
    HCOOCH$_3$ (d) calculated at ionization rates of $10^{-17}$ s$^{-1}$ (solid
    line), $10^{-16}$ s$^{-1}$ (dotted line), $10^{-15}$ s$^{-1}$ (dashed
    line), and $10^{-14}$ s$^{-1}$ (dot-dashed line).
}
  \label{f8}
\end{figure}

\begin{figure}[htb!]
  \gridline{
  \fig{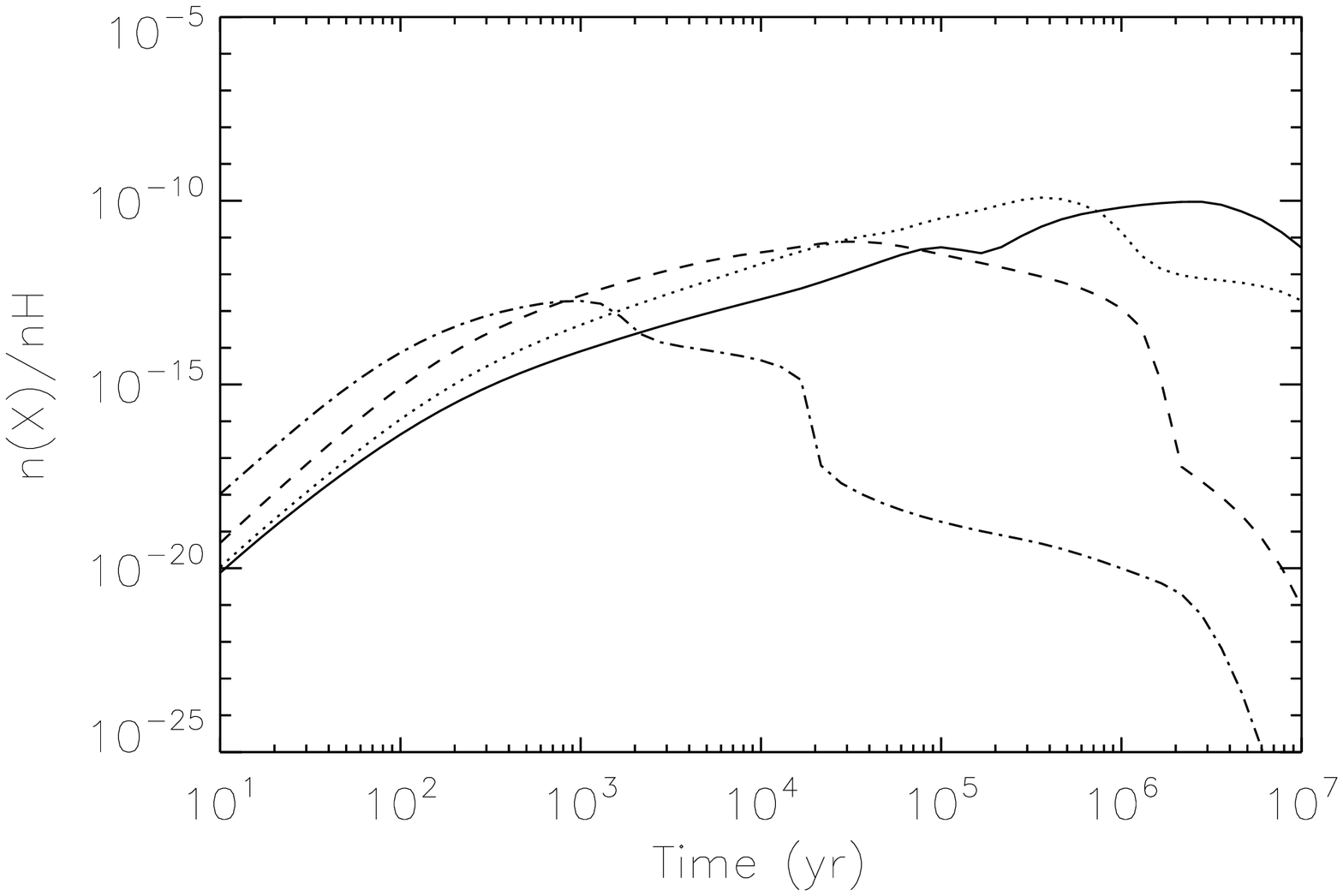}{0.5\textwidth}{(a)}
  \fig{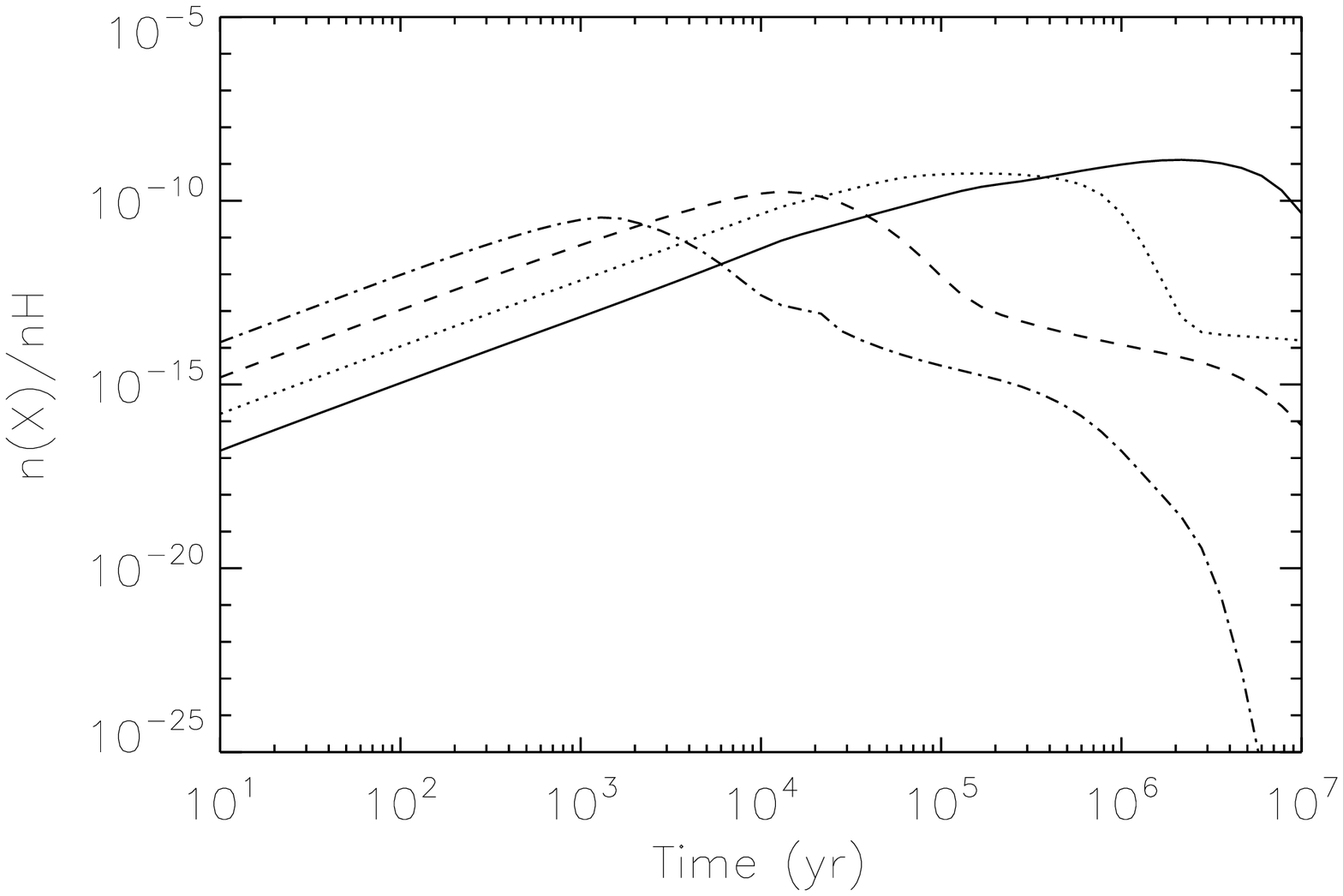}{0.5\textwidth}{(b)}
  }
  \gridline{
  \fig{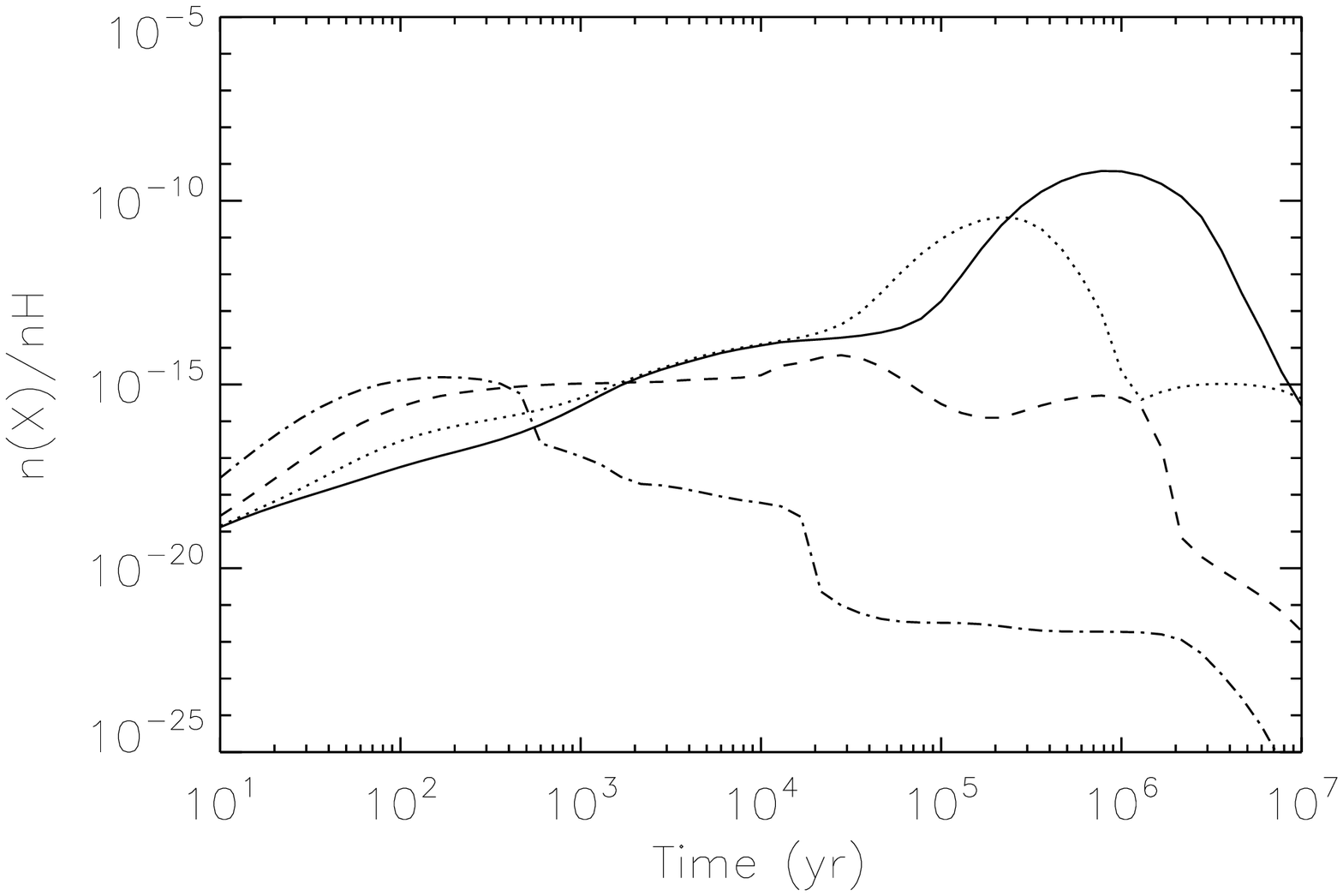}{0.5\textwidth}{(c)}
  \fig{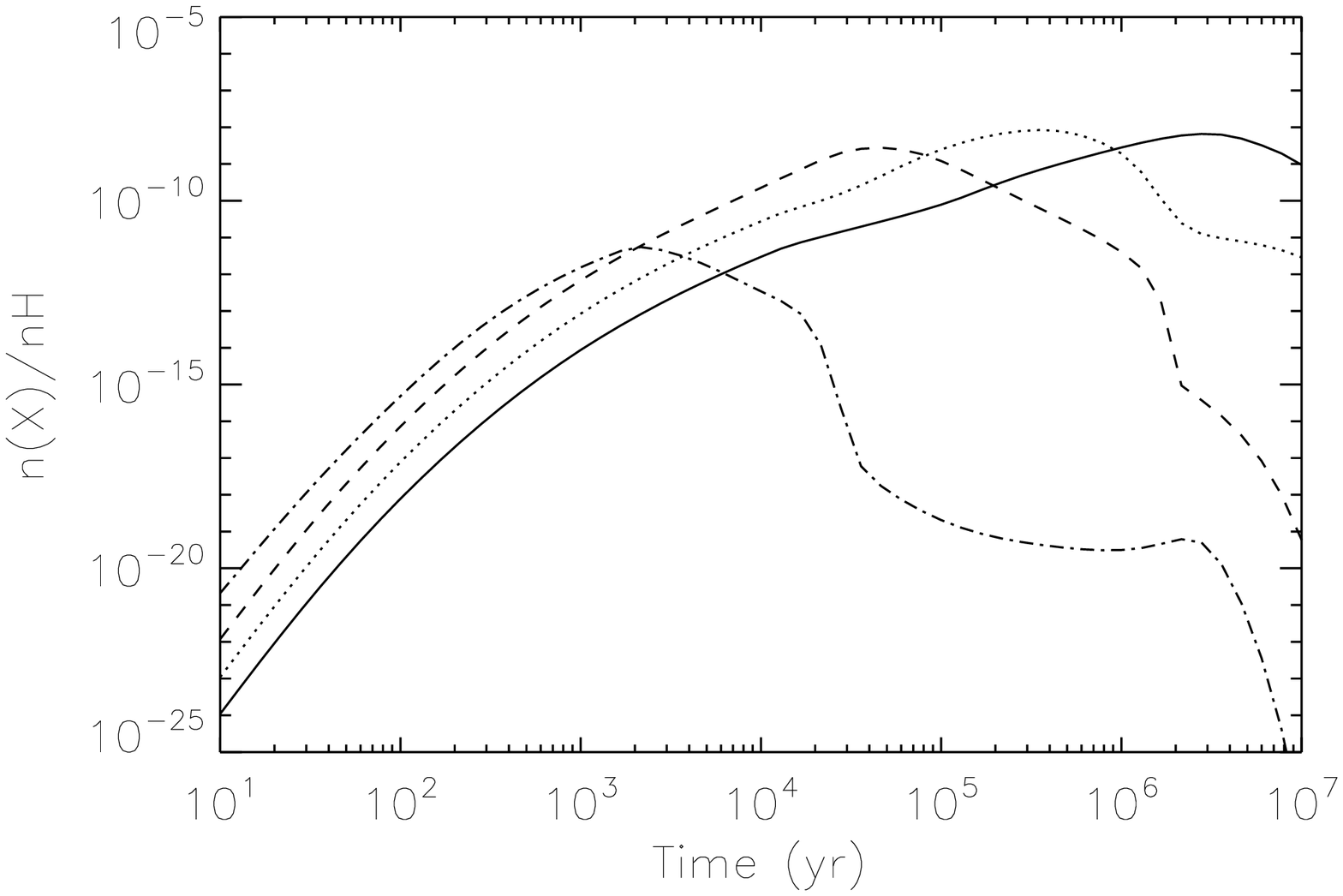}{0.5\textwidth}{(d)}
  }
  \caption{
    Calculated grain-surface abundances of HOCO (a) NO$_2$ (b) HC$_2$O (c),
    and HCOOCH$_3$ (d) calculated at ionization rates of $10^{-17}$ s$^{-1}$
    (solid line), $10^{-16}$ s$^{-1}$ (dotted line), $10^{-15}$ s$^{-1}$
    (dashed line), and $10^{-14}$ s$^{-1}$ (dot-dashed line).
}
  \label{f9}
\end{figure}

\begin{figure}[htb!]
  \gridline{
  \fig{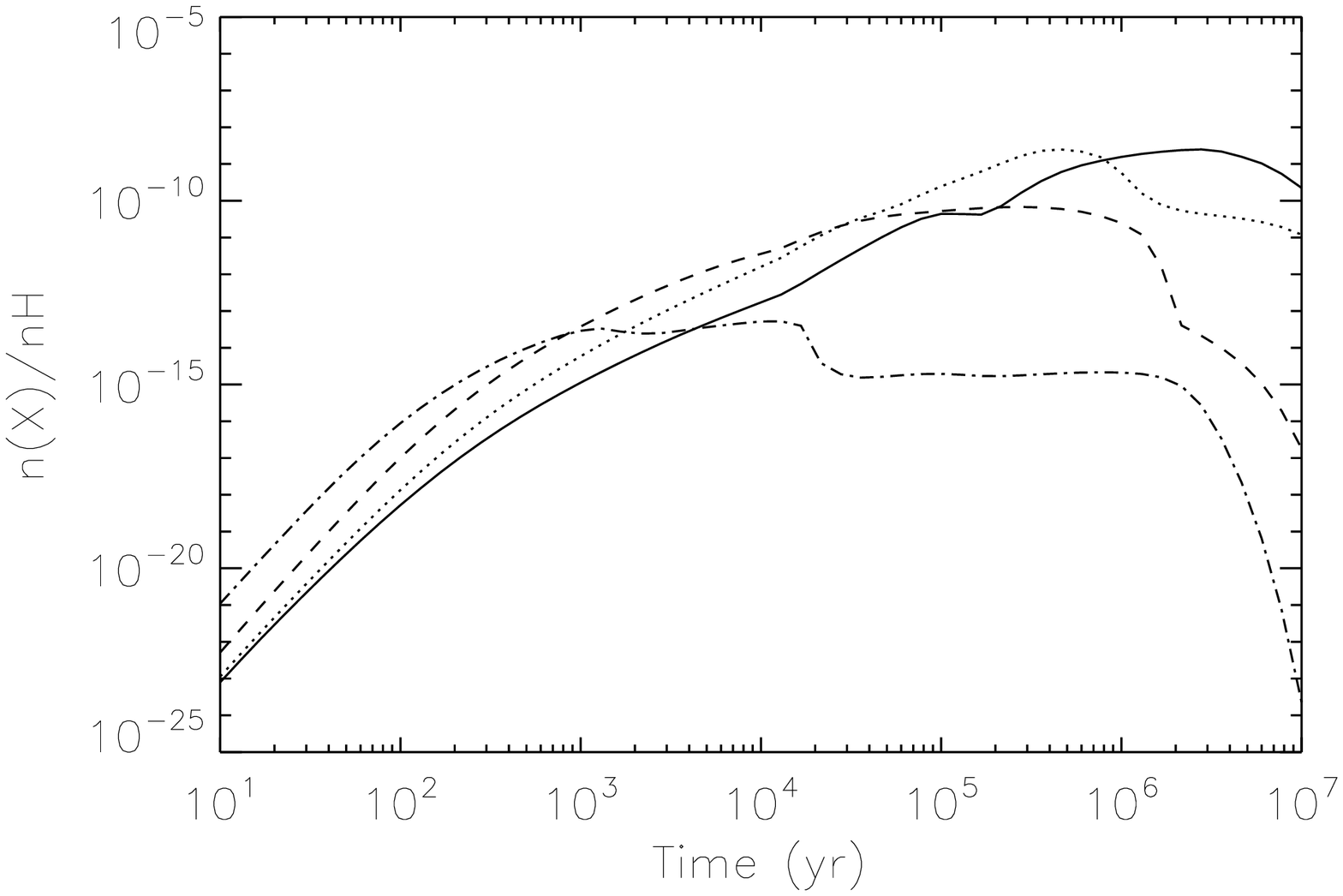}{0.5\textwidth}{(a)}
  \fig{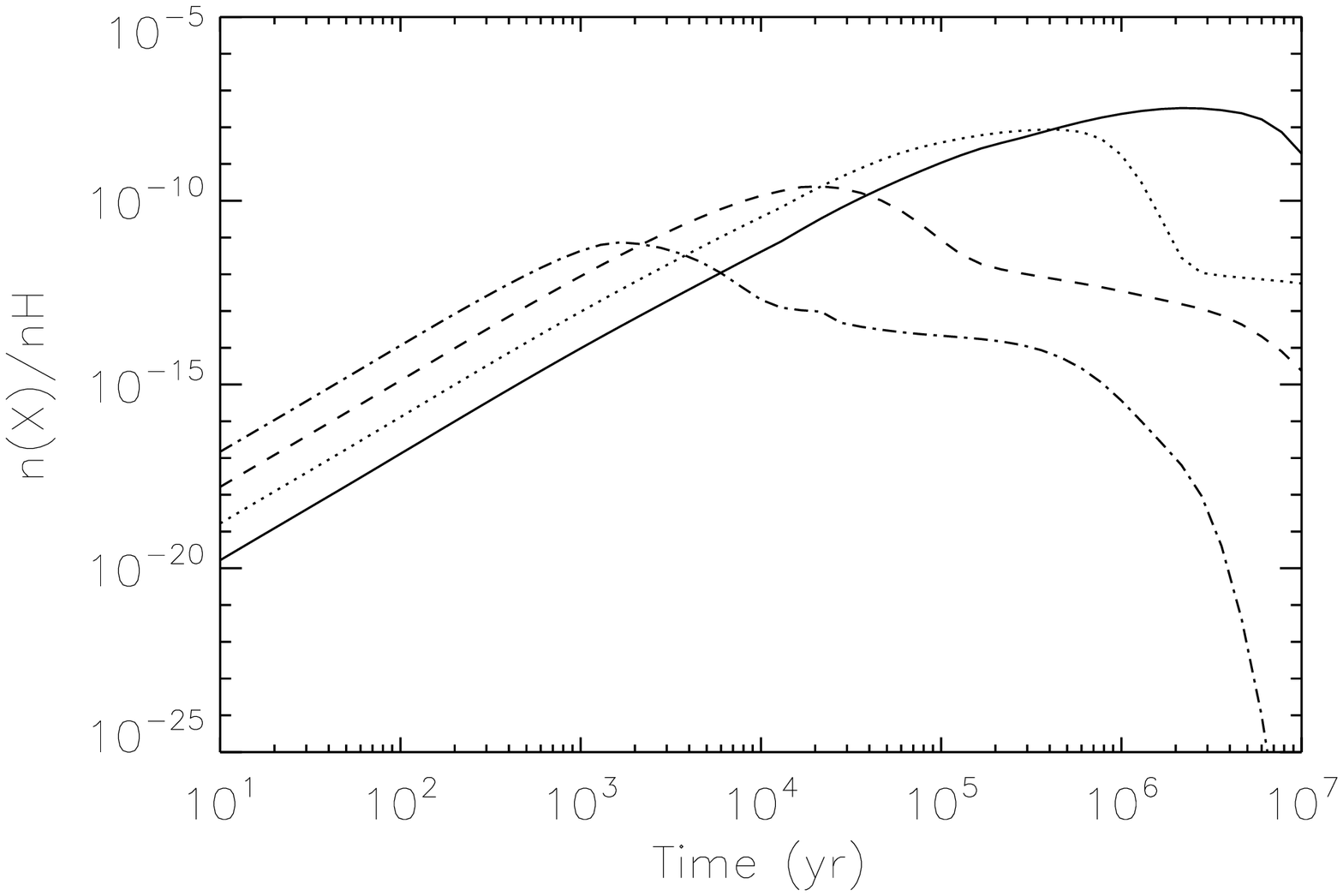}{0.5\textwidth}{(b)}
  }
  \gridline{
  \fig{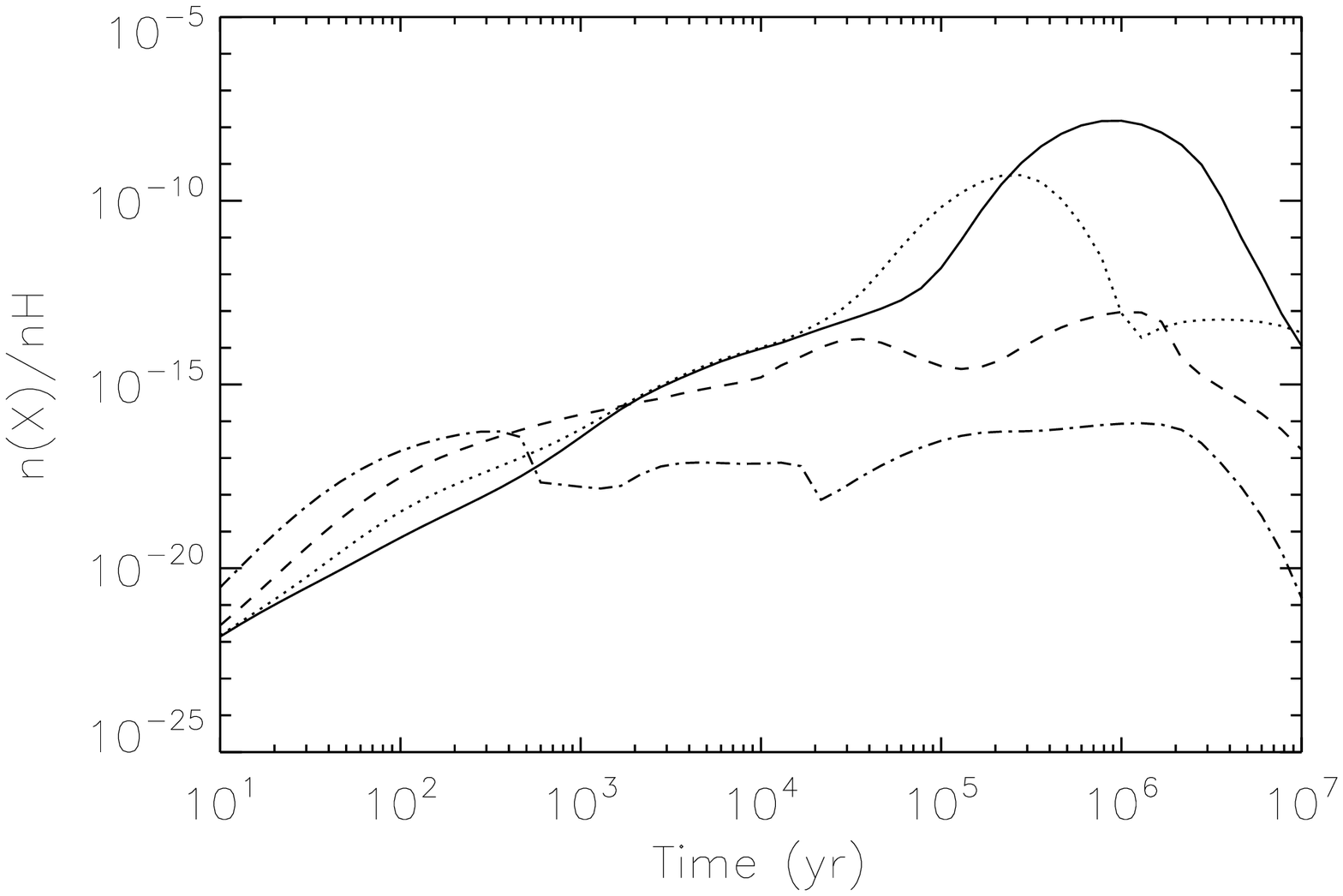}{0.5\textwidth}{(c)}
  \fig{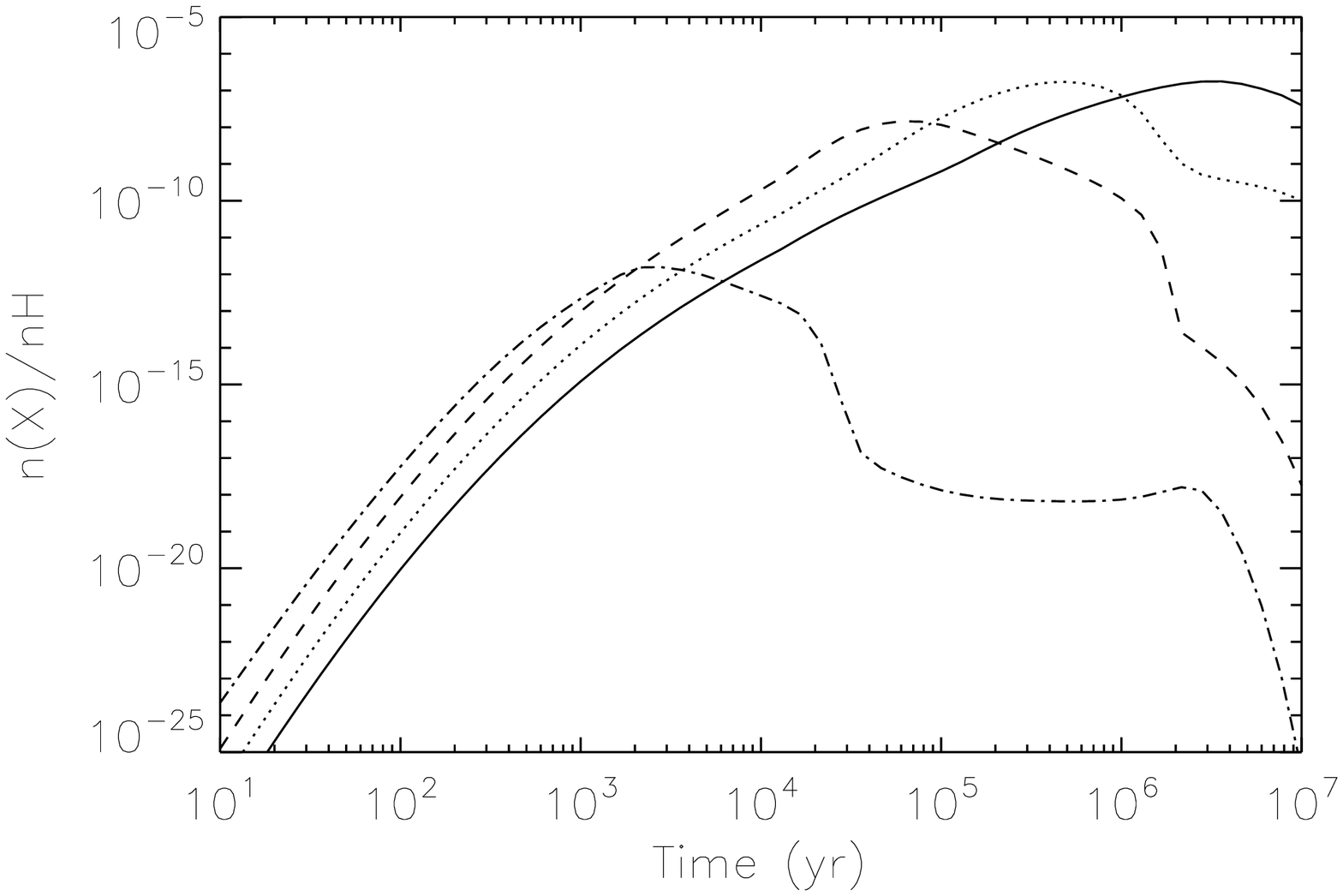}{0.5\textwidth}{(d)}
  }
  \caption{
    Calculated bulk-ice abundances of HOCO (a) NO$_2$ (b) HC$_2$O (c), and
    HCOOCH$_3$ (d) calculated at ionization rates of $10^{-17}$ s$^{-1}$ (solid
    line), $10^{-16}$ s$^{-1}$ (dotted line), $10^{-15}$ s$^{-1}$ (dashed
    line), and $10^{-14}$ s$^{-1}$ (dot-dashed line).
}
  \label{f10}
\end{figure}

\FloatBarrier

Additional simulations were run in order to examine the effect of the new radiation chemistry at high $\zeta$. As mentioned in \S \ref{sec:model} - and shown in Table \ref{tab:parameters} - we assume that the simulated hypothetical sources are physically identical to TMC-1  except for having higher ionization rates. The results from these model runs for HOCO, NO$_2$, HC$_2$O, and HCOOCH$_3$ are depicted in Figs. \ref{f8}-\ref{f10}, which show the gas, surface, and bulk abundances, respectively. 

As one can see from a comparison of Figs. \ref{f8}-\ref{f10}, several trends emerge as the ionization rate changes. First, since, as previously demonstrated, the abundances of HOCO, NO$_2$, HC$_2$O, and HCOOCH$_3$ are enhanced due to radiochemical processes, it is reasonable that their abundances should tend to increase with increasing $\zeta$. This effect is most obvious at very early times before $\sim10^3$ yr, with the correlation between the two clearly observable in Figs. \ref{f8}-\ref{f10}. At intermediate times however, between $\sim10^3$-$10^6$ yr, the relationship between abundance and $\zeta$ begins to break down, particularly in the gas phase. Generally, we find that the higher the ionization rate, the faster the peak abundance is reached, and the lower the peak value - a trend that can most easily be seen in Figs. \ref{f9} and \ref{f10}, which show the surface and bulk abundances, respectively. After $\sim 10^6$ yr, an anti-correlation between $\zeta$ and abundance emerges for most of the species shown. The reasons for this behavior are complex, but are driven in part by (a) the increased radiolytic destruction of surface and bulk species into more weakly bound fragments, and (b) the greatly increased gas-phase abundances of ions such as H$^+$ and C$^+$, reactions with which further reduce the abundance of the neutral species considered here.

\FloatBarrier

\section{Conclusions} \label{sec:conclusion}

We have utilized the theory described in \citet{shingledecker_general_2017} in an initial attempt to incorporate radiation chemistry into an existing chemical network. Simulations of the cold core TMC-1 were run, both with and without the new cosmic ray-induced reactions. We also modeled several hypothetical sources which were physically identical to TMC-1 other than having enhanced ionization rates. The major results of the simulations described in this work are the following:

\begin{itemize}
  \item Radiation chemistry can result in substantially enhanced abundances
    in all three model phases for a variety of species, including COMs.
  \item These enhancements in abundance occur mainly as a result of reactions
    involving suprathermal species formed from the radiolytic dissociation of simple
    ice mantle constituents.
  \item Even under cold core conditions, these suprathermal species
    can react quickly by a variety of  mechanisms, including insertion,
    which we found to be particularly important in increasing the abundance of
    COMs.
  \item We predict that HOCO, and perhaps NO$_2$, could be observable in TMC-1,
    given a sufficiently deep search.
  \item The addition of radiation chemistry substantially improves agreement
    between calculated and observed abundances of HC$_2$O.
  \item For the neutral species considered here, ionization rates of $10^{-16}$ s$^{-1}$ or higher generally resulted in reduced abundances in all model phases at times greater than $\sim 10^3$ yr.
\end{itemize}

It should be emphasized that these results, while promising, are necessarily
preliminary in nature, given the novelty of incorporating radiation chemistry
into astrochemical models.  More work is needed to better characterize both (a)
cosmic ray-induced radiolysis and chemistry and (b) secondary effects such as
the non-thermal desorption of grain species triggered by cosmic ray
bombardment.  These non-thermal desorption mechanisms, such as sputtering,
desorption induced by electronic transitions (DIET), electron stimulated ion
desorption (ESID), and Auger stimulated ion desorption (ASID)
\citep{ribeiro_non-thermal_2015} are particularly promising since they could
provide a means of enriching gas-phase abundances at low temperatures, and are
therefore a natural complement to the non-thermal chemistry described here.

As we have demonstrated in this work, the addition of cosmic ray-driven
solid-phase reactions can improve existing astrochemical models in a number of
significant ways. First, the addition of this non-thermal chemistry increases
the  realism of models, since cosmic ray bombardment of
ice mantles certainly occurs in the ISM.  Moreover, a consideration of
solid-phase radiation chemistry both helps to explain how COMs like methyl
formate could efficiently form in cold cores
\citep{balucani_formation_2015,chang_unified_2016,vasyunin_reactive_2013}, and improves the agreement
between calculated and observational abundances for HC$_2$O. Cosmic ray-driven
ice chemistry is thus attractive as a component of future astrochemical
modeling studies. 

E. H. wishes to thank the National Science Foundation for supporting
the astrochemistry program at the University of Virginia through grant AST 15 - 14844. C. N. S. thanks V. Wakelam for use of the \texttt{NAUTILUS-1.1} code. This research has made use of NASA's Astrophysics Data System Bibliographic Services

\FloatBarrier
\appendix
\section{Radiolysis Reactions}

\startlongtable
\label{tab:radiolysis}
\begin{deluxetable}{llccccc}
  \tablecaption{New solid-phase radiolysis processes}
  \tablehead{
    \colhead{Number} & \colhead{Process} & \colhead{} & \colhead{} & \colhead{$f_\mathrm{br}$} & \colhead{$G$-value} & \colhead{Type}
  }
  \startdata
  \multicolumn{7}{c}{H$_2$O} \\
  1 & $\mathrm{H_2O}\leadsto \mathrm{O^*}+\mathrm{H_2^*}$ & & & $0.500$  & $3.704$ & $\RN{1}$ \\
  2 & $\mathrm{H_2O}\leadsto \mathrm{OH^*}+\mathrm{H^*}$ & & & $0.500$  & $3.704$ & $\RN{1}$ \\
  3 & $\mathrm{H_2O}\leadsto \mathrm{OH}+\mathrm{H}$ & & & $1.000$  & $1.747$ & $\RN{2}$ \\
  4 & $\mathrm{H_2O}\leadsto \mathrm{H_2O}$ & & & $1.000$  & $1.747$ & $\RN{3}$ \\
  \hline 
  \multicolumn{7}{c}{O$_2$} \\
  5 & $\mathrm{O_2}\leadsto \mathrm{O^*}+\mathrm{O^*}$ & & & $1.000$  & $3.704$ & $\RN{1}$ \\
  6 & $\mathrm{O_2}\leadsto \mathrm{O}+\mathrm{O}$ & & & $1.000$  & $2.138$ & $\RN{2}$ \\
  7 & $\mathrm{O_2}\leadsto \mathrm{O_2^*}$ & & & $1.000$  & $2.138$ & $\RN{3}$ \\
  \hline 
  \multicolumn{7}{c}{O$_3$} \\
  8 & $\mathrm{O_3}\leadsto \mathrm{O_2^*}+\mathrm{O^*}$ & & & $1.000$  & $3.704$ & $\RN{1}$ \\
  9 & $\mathrm{O_3}\leadsto \mathrm{O_2}+\mathrm{O}$ & & & $1.000$  & $4.059$ & $\RN{2}$ \\
  10 & $\mathrm{O_3}\leadsto \mathrm{O_3^*}$ & & & $1.000$  & $4.059$ & $\RN{3}$ \\
  \hline 
  \multicolumn{7}{c}{CO} \\
  11 & $\mathrm{CO}\leadsto \mathrm{C^*}+\mathrm{O^*}$ & & & $1.000$  & $3.704$ & $\RN{1}$ \\
  12 & $\mathrm{CO}\leadsto \mathrm{C}+\mathrm{O}$ & & & $1.000$  & $1.269$ & $\RN{2}$ \\
  13 & $\mathrm{CO}\leadsto \mathrm{CO^*}$ & & & $1.000$  & $1.269$ & $\RN{3}$ \\
  \hline 
  \multicolumn{7}{c}{CO$_2$} \\
  14 & $\mathrm{CO_2}\leadsto \mathrm{CO^*}+\mathrm{O^*}$ & & & $1.000$  & $3.704$ & $\RN{1}$ \\
  15 & $\mathrm{CO_2}\leadsto \mathrm{CO}+\mathrm{O}$ & & & $1.000$  & $1.249$ & $\RN{2}$ \\
  16 & $\mathrm{CO_2}\leadsto \mathrm{CO_2^*}$ & & & $1.000$  & $1.249$ & $\RN{3}$ \\
  \hline 
  \multicolumn{7}{c}{NO} \\
  17 & $\mathrm{NO}\leadsto \mathrm{N^*}+\mathrm{O^*}$ & & & $1.000$  & $3.704$ & $\RN{1}$ \\
  18 & $\mathrm{NO}\leadsto \mathrm{N}+\mathrm{O}$ & & & $1.000$  & $1.922$ & $\RN{2}$ \\
  19 & $\mathrm{NO}\leadsto \mathrm{NO^*}$ & & & $1.000$  & $1.922$ & $\RN{3}$ \\
  \hline 
  \multicolumn{7}{c}{NO$_2$} \\
  20 & $\mathrm{NO}\leadsto \mathrm{NO^*}+\mathrm{O^*}$ & & & $1.000$  & $3.704$ & $\RN{1}$ \\
  21 & $\mathrm{NO}\leadsto \mathrm{NO}+\mathrm{O}$ & & & $1.000$  & $1.207$ & $\RN{2}$ \\
  22 & $\mathrm{NO}\leadsto \mathrm{NO_2^*}$ & & & $1.000$  & $1.207$ & $\RN{3}$ \\
  \hline 
  \multicolumn{7}{c}{O$_2$H} \\
  23 & $\mathrm{O_2H}\leadsto \mathrm{OH^*}+\mathrm{O^*}$ & & & $1.000$  & $3.704$ & $\RN{1}$ \\
  24 & $\mathrm{O_2H}\leadsto \mathrm{OH}+\mathrm{O}$ & & & $1.000$  & $3.714$ & $\RN{2}$ \\
  25 & $\mathrm{O_2H}\leadsto \mathrm{O_2H^*}$ & & & $1.000$  & $3.714$ & $\RN{3}$ \\
  \hline 
  \multicolumn{7}{c}{H$_2$O$_2$} \\
  26 & $\mathrm{H_2O_2}\leadsto \mathrm{OH^*}+\mathrm{OH^*}$ & & & $0.500$  & $3.704$ & $\RN{1}$ \\
  27 & $\mathrm{H_2O_2}\leadsto \mathrm{O^*}+\mathrm{H_2O^*}$ & & & $0.500$  & $3.704$ & $\RN{1}$ \\
  28 & $\mathrm{H_2O_2}\leadsto \mathrm{OH}+\mathrm{OH^*}$ & & & $1.000$  & $2.296$ & $\RN{2}$ \\
  \hline 
  \multicolumn{7}{c}{NH$_3$} \\
  29 & $\mathrm{NH_3}\leadsto \mathrm{H^*}+\mathrm{NH_2^*}$ & & & $0.500$  & $3.704$ & $\RN{1}$ \\
  30 & $\mathrm{NH_3}\leadsto \mathrm{H_2^*}+\mathrm{NH^*}$ & & & $0.500$  & $3.704$ & $\RN{1}$ \\
  31 & $\mathrm{NH_3}\leadsto \mathrm{H}+\mathrm{NH_2}$ & & & $1.000$  & $2.721$ & $\RN{2}$ \\
  32 & $\mathrm{NH_3}\leadsto \mathrm{NH_3^*}$ & & & $1.000$  & $2.721$ & $\RN{3}$ \\
  \hline 
  \multicolumn{7}{c}{CH$_4$} \\
  33 & $\mathrm{CH_4}\leadsto \mathrm{H^*}+\mathrm{CH_3^*}$ & & & $0.500$  & $3.704$ & $\RN{1}$ \\
  34 & $\mathrm{CH_4}\leadsto \mathrm{H_2}+\mathrm{CH_2^*}$ & & & $0.500$  & $3.704$ & $\RN{1}$\tablenotemark{a} \\
  35 & $\mathrm{CH_4}\leadsto \mathrm{H}+\mathrm{CH_3}$ & & & $1.000$  & $1.505$ & $\RN{2}$ \\
  36 & $\mathrm{CH_4}\leadsto \mathrm{CH_4^*}$ & & & $1.000$  & $1.505$ & $\RN{3}$ \\
\hline 
\multicolumn{7}{c}{H$_2$CO} \\
  37 & $\mathrm{H_2CO}\leadsto \mathrm{H^*}+\mathrm{HCO^*}$ & & & $1.000$  & $3.704$ & $\RN{1}$ \\
  38 & $\mathrm{H_2CO}\leadsto \mathrm{H}+\mathrm{HCO}$ & & & $1.000$  & $2.910$ & $\RN{2}$ \\
  39 & $\mathrm{H_2CO}\leadsto \mathrm{H_2CO^*}$ & & & $1.000$  & $2.910$ & $\RN{1}$ \\
\hline 
\multicolumn{7}{c}{CH$_3$OH} \\
  40 & $\mathrm{CH_3OH}\leadsto \mathrm{H^*}+\mathrm{CH_3O^*}$ & & & $0.333$  & $3.704$ & $\RN{1}$ \\
  41 & $\mathrm{CH_3OH}\leadsto \mathrm{H^*}+\mathrm{CH_2OH^*}$ & & & $0.333$  & $3.704$ & $\RN{1}$ \\
  42 & $\mathrm{CH_3OH}\leadsto \mathrm{OH^*}+\mathrm{CH_3^*}$ & & & $0.333$  & $3.704$ & $\RN{1}$ \\
  43 & $\mathrm{CH_3OH}\leadsto \mathrm{H}+\mathrm{CH_3O}$ & & & $0.333$  & $1.571$ & $\RN{2}$ \\
  44 & $\mathrm{CH_3OH}\leadsto \mathrm{H}+\mathrm{CH_2OH}$ & & & $0.333$  & $1.571$ & $\RN{2}$ \\
  45 & $\mathrm{CH_3OH}\leadsto \mathrm{OH}+\mathrm{CH_3}$ & & & $0.333$  & $1.571$ & $\RN{2}$ \\
  46 & $\mathrm{CH_3OH}\leadsto \mathrm{CH_3OH^*}$ & & & $1.000$  & $1.571$ & $\RN{3}$ \\
\hline 
\multicolumn{7}{c}{CH$_3$COCH$_3$} \\
  47 & $\mathrm{CH_3COCH_3}\leadsto \mathrm{CH_3^*}+\mathrm{CH_3CO^*}$ & & & $1.000$  & $3.704$ & $\RN{1}$\tablenotemark{b} \\
  48 & $\mathrm{CH_3COCH_3}\leadsto \mathrm{CH_3}+\mathrm{CH_3CO}$ & & & $1.000$  & $4.020$ & $\RN{2}$ \\
  49 & $\mathrm{CH_3COCH_3}\leadsto \mathrm{CH_3COCH_3^*}$ & & & $1.000$  & $4.020$ & $\RN{3}$ \\
\enddata
  \tablenotetext{a}{\citet{bergantini_mechanistical_2018}}
  \tablenotetext{b}{\citet{hudson_radiation_2017}}
\end{deluxetable}

\FloatBarrier

\section{Class 2 Reactions}

%\startlongtable
\begin{deluxetable}{llccccr}[ht]
  \label{tab:c2novel}
  \tablecaption{New Class 2 reactions involving suprathermal species.}
  \tablehead{
    \colhead{Number} & \colhead{Reaction} & \colhead{} & \colhead{} & \colhead{} & \colhead{$f_\mathrm{br}$} & \colhead{Source}
  }
  \startdata
  \multicolumn{7}{c}{C$^*$} \\
  50 & $\mathrm{C^*} + \mathrm{H_2O} \rightarrow \mathrm{CH} + \mathrm{OH}$ & & & & 1.0 & \citet{mayer_computation_1967} \\
  51 & $\mathrm{C^*} + \mathrm{CO} \rightarrow \mathrm{CCO}$ & & & & 1.0 & \citet{husain_reactions_1971} \\
  52 & $\mathrm{C^*} + \mathrm{CH_3OH} \rightarrow \mathrm{CH_3CHO}$ & & & & 0.5 & \citet{shannon_fast_2014} \\
  53 & $\mathrm{C^*} + \mathrm{CH_3OH} \rightarrow \mathrm{CH_3} + \mathrm{HCO}$ & & & & 0.5 & \citet{shannon_fast_2014} \\
  \hline
  \multicolumn{7}{c}{O$^*$} \\
  54 & $\mathrm{O^*} + \mathrm{CH_4} \rightarrow \mathrm{CH_3OH}$ & & & & 0.65 & \citet{bergner_methanol_2017} \\
  55 & $\mathrm{O^*} + \mathrm{CH_4} \rightarrow \mathrm{H_2CO} + \mathrm{H_2}$ & & & & 0.35 & \citet{bergner_methanol_2017} \\
  56 & $\mathrm{O^*} + \mathrm{CH_3OH} \rightarrow \mathrm{CH_3} + \mathrm{HCO}$ & & & & 1.0 & \citet{matsumi_isotopic_1994} \\
  57 & $\mathrm{O^*} + \mathrm{NO} \rightarrow \mathrm{NO_2}$ & & & & 1.0 & \citet{atkinson_evaluated_2004} \\
  \hline
  \multicolumn{7}{c}{CH$_2^*$} \\
  58 & $\mathrm{CH_2^*} + \mathrm{CH_3OH} \rightarrow \mathrm{CH_3CH_2OH}$ & & & & 0.5 & \citet{bergantini_mechanistical_2018} \\
  59 & $\mathrm{CH_2^*} + \mathrm{CH_3OH} \rightarrow \mathrm{CH_3OCH_3}$ & & & & 0.5 & \citet{bergantini_mechanistical_2018} \\
  \enddata
\end{deluxetable}

\FloatBarrier
\section{New HOCO Reactions}

\startlongtable
\begin{deluxetable}{llcccr}
  \label{tab:hoco}
  \tablecaption{New gas-phase HOCO destruction reactions}
  \tablehead{
    \colhead{Number} & \colhead{Reaction} & \colhead{$\alpha$} & \colhead{$\beta$} & \colhead{$\gamma$} & \colhead{Source} \\
  }
  \startdata
  \multicolumn{6}{c}{Neutral-Neutral\tablenotemark{a}} \\
  %Reaction & $\alpha$ & $\beta$ & $\gamma$ & Source \\
  &  & s$^{-1}$ & K &  \\
  \hline \\
  60 & $\mathrm{HOCO}+\mathrm{Cl}\rightarrow \mathrm{HCl}+\mathrm{CO_2}$ & $4.800\times 10^{-11}$ & 0.000 & 0.000 & \citet{li_rate_2000} \\
  61 & $\mathrm{HOCO}+\mathrm{O_2}\rightarrow \mathrm{O_2H}+\mathrm{CO_2}$ & $1.900\times 10^{-12}$ & 0.000 & 0.000 & \citet{poggi_ab_2004} \\
  62 & $\mathrm{HOCO}+\mathrm{NO}\rightarrow \mathrm{HNO}+\mathrm{CO_2}$ & $2.450\times 10^{-12}$ & 0.000 & 0.000 & \citet{poggi_ab_2004} \\
  63 & $\mathrm{HOCO}+\mathrm{O}\rightarrow \mathrm{OH}+\mathrm{CO_2}$ & $1.440\times 10^{-11}$ & 0.000 & 0.000 & \citet{yu_quantum_2007} \\
  64 & $\mathrm{HOCO}+\mathrm{OH}\rightarrow \mathrm{H_2O}+\mathrm{CO_2}$ & $1.030\times 10^{-11}$ & 0.000 & 0.000 & \citet{yu_direct_2005} \\
  65 & $\mathrm{HOCO}+\mathrm{CH_3}\rightarrow \mathrm{H_2O}+\mathrm{H2C_2O}$ & $5.800\times 10^{-11}$  & 0.000 & 0.000 & \citet{yu_theoretical_2009} \\
  \hline
  \multicolumn{6}{c}{Ion-Neutral\tablenotemark{b}} \\
  &  & $f_\mathrm{br}$ & cm$^3$ s$^{-1}$ &  \\
  \hline
  66 & $\mathrm{HOCO}+\mathrm{H^+}\rightarrow \mathrm{HOCO^+}+\mathrm{H}$ & $1.000$ & $5.049\times 10^{-9}$ & 9.438 & See Text \\
  67 & $\mathrm{HOCO}+\mathrm{H_3^+}\rightarrow \mathrm{HOCO^+}+\mathrm{H}+\mathrm{H_2}$ & $1.000$ & $2.978\times 10^{-9}$ & 9.438 & See Text \\
  68 & $\mathrm{HOCO}+\mathrm{He^+}\rightarrow \mathrm{HOCO^+}+\mathrm{He}$ & $1.000$ & $3.609\times 10^{-9}$ & 9.438 & See Text \\
  69 & $\mathrm{HOCO}+\mathrm{C^+}\rightarrow \mathrm{HOCO^+}+\mathrm{C}$ & $1.000$ & $1.623\times 10^{-9}$ & 9.438 & See Text \\
  \enddata
  \tablenotetext{a}{See Eq. \eqref{ak_formula}}
  \tablenotetext{b}{See \citet{woon_quantum_2009}}
\end{deluxetable}

\bibliography{bibliography}
\bibliographystyle{aasjournal}

\end{document}